\documentclass[%
 reprint,
 citeautoscript,
 amsmath,amssymb,
 aps,
 prb,
]{revtex4-2}

\usepackage{graphicx}
\usepackage{dcolumn}
\usepackage{bm}
\usepackage{xcolor}
\usepackage{hyperref}
\hypersetup{
    colorlinks=true,
    linkcolor=blue,
    citecolor=blue,
    filecolor=blue,      
    urlcolor=blue,
    }
\usepackage[T1]{fontenc}

\newcommand{\FS}[1]{\left\langle #1 \right\rangle_{\scriptscriptstyle\mathrm{FS}}}
\newcommand{\FSplus}[1]{\left\langle #1 \right\rangle_{+}}
\newcommand{\FSminus}[1]{\left\langle #1 \right\rangle_{-}}
 
\newcommand{\kb}{k_\mathrm{B}}
\newcommand{\tc}{T_\mathrm{c0}}
\newcommand{\tcs}{T_\mathrm{s}}

\newcommand{\RR}{\mathbf{R}}
\newcommand{\vF}{\mathbf{v}_\mathrm{F}}
\newcommand{\pF}{\mathbf{p}_\mathrm{F}}
\newcommand{\ps}{\mathbf{p}_\mathrm{s}}

\newcommand{\NF}{\mathcal{N}_\mathrm{F}}

\newcommand{\fe}{f_\mathrm{e}}
\newcommand{\vfout}{\vF^{\mathrm{out}}}
\newcommand{\vfin}{\vF^{\mathrm{in}}}

\newcommand{\xx}{\emph{\fontfamily{cmss}\selectfont x}}
\newcommand{\xco}{x}

\begin{document}

\title{Self-consistent theory of current injection into $d$ and $d+is$ superconductors}

\author{Kevin Marc Seja}
 \affiliation{Department of Microtechnology and Nanoscience - MC2,
Chalmers University of Technology,
SE-41296 G\"oteborg, Sweden}
\author{Tomas L\"ofwander}
 \affiliation{Department of Microtechnology and Nanoscience - MC2,
Chalmers University of Technology,
SE-41296 G\"oteborg, Sweden}

\date{\today}

\begin{abstract}

We present results for the steady-state nonlinear response of a $d_{x^2-y^2}$ superconducting film connected to normal-metal reservoirs under voltage bias, allowing for a subdominant $s$-wave component appearing near the interfaces. Our investigation is based on a current-conserving theory that self-consistently includes the non-equilibrium distribution functions, charge imbalance, and the voltage-dependencies of order parameters and scalar impurity self-energies. For a pure $d$-wave superconductor with [110] orientation of the interfaces to the contacts, the conductance contains a zero-bias peak reflecting the large density of zero-energy interface Andreev bound states. Including a subdominant $s$-wave pairing channel, it is in equilibrium energetically favorable for an $s$-wave order parameter component $\Delta_\mathrm{s}$ to appear near the interfaces in the time-reversal symmetry breaking combination $d+is$. The Andreev states then shift to finite energies in the density of states. Under voltage bias, we find that the non-equilibrium distribution in the contact area causes a rapid suppression of the $s$-wave component to zero as the voltage $eV\rightarrow\Delta_\mathrm{s}$. The resulting spectral rearrangements and voltage-dependent scattering amplitudes lead to a pronounced non-thermally broadened split of the zero-bias conductance peak that is not seen in a non-selfconsistent Landauer-B\"uttiker scattering approach. 

\end{abstract}

\maketitle

\section{Introduction}\label{sec:introduction}
Tunneling spectroscopy was one of the early experimental methods to extract information about the energy gap as predicted by the Bardeen-Cooper-Schrieffer theory of superconductivity \cite{Giaever1960,Bardeen1961,Schrieffer:Book}. Following the generalization to point-contact Andreev reflection spectroscopy by Blonder-Tinkham-Klapwijk (BTK) \cite{Blonder1982Apr}, where higher order Andreev processes are taken into account for more transparent point contacts, it has also served as a tool to extract the spin-polarization of ferromagnets \cite{Merservey1994,Soulen1998,Upadhyay1998} and probe the symmetry of the order parameter of unconventional superconductors \cite{Geerk1988,Hu1994,Tanaka1995Apr,Covington1997Jul,Fogelstrom1997Jul,kashiwaya_tunnelling_2000,lofwander_andreev_2001}. A challenge is that the complicated physics at the contact may play a crucial role. This is particularly important when it comes to applying the method to unconventional superconductors, where the superconducting order parameter may be sensitive to normal reflection at the contact. The pair breaking is accompanied by the formation of interface Andreev states, that have also been taken as fingerprints of the symmetry of the order parameter \cite{kashiwaya_tunnelling_2000,Tsuei2000Oct,lofwander_andreev_2001}.

For the high-temperature superconductors with $d_{x^2-y^2}$ symmetry of the order parameter, tunneling and Andreev reflection spectroscopy typically show a large zero-bias conductance peak (ZBCP) \cite{Geerk1988,Hu1994,Tanaka1995Apr,Covington1997Jul,Fogelstrom1997Jul,kashiwaya_tunnelling_2000,lofwander_andreev_2001}. The ZBCP appears due to the large spectral weight of zero-energy Andreev bound states at surfaces oriented 45$^\circ$ relative to the crystal $ab$-axes, also denoted [110] surfaces. These Andreev states are formed due to the different signs of the $d_{x^2-y^2}$ order parameter around the Fermi surface \cite{Hu1994}. In a magnetic field, these Andreev states are shifted to finite energies by the screening supercurrents, leading to a field-dependent split of the ZBCP, an effect that is well established experimentally \cite{Covington1997Jul,Fogelstrom1997Jul}. A spontaneous split of the ZBCP in the absence of magnetic field has been observed in several experiments, but not all \cite{Covington1997Jul,Krupke1999,Wei1998,Dagan2001,Elhalel2007Mar,Ngai2010Aug}. Such a split indicates the possibility of a subdominant component of the order parameter of either $d_{xy}$ or $s$ symmetry, combined with the main $d_{x^2-y^2}$ component in a time-reversal symmetry breaking $d_{x^2-y^2}+id_{xy}$ or $d_{x^2-y^2}+is$ state, at least at the surface \cite{Matsumoto1995,Fogelstrom1997Jul,Sigrist1998}. Other experiments either support time-reversal symmetry breaking or give severe size constraints on the subdominant order parameter \cite{Carmi2000,Neils2002,Gustafsson2013,Kirtley2006,Saadaoui2011}. One possible explanation for the absence of the split of the ZBCP that has been proposed recently is the formation of an inhomogeneous state at the edge with spontaneous circulating currents below a relatively large transition temperature $T^*\approx 0.18\tc$ \cite{Hakansson2015,Holmvall2018}. This time-reversal symmetry broken state is  referred to as phase crystal \cite{Holmvall2020} and is robust also in presence of strong correlations and Anderson disorder \cite{Chakraborty2022}. If this state is formed it also suppresses the nucleation of the subdominant component if the criticial temperature for the formation of subdominant order, $\tcs$, is too low. On the other hand, if $\tcs$ is sufficiently large, it will instead prevent the phase crystal to be formed \cite{Hakansson2015}. We note that the large spectral weight of Andreev surface states could also lead to other instabilities, such as spontaneous magnetization \cite{Honerkamp2000,Potter2014}. In conclusion, the breaking of time-reversal symmetry remains a topic of high interest in this research field. 

\begin{figure}
    \centering
    \includegraphics[width=\columnwidth]{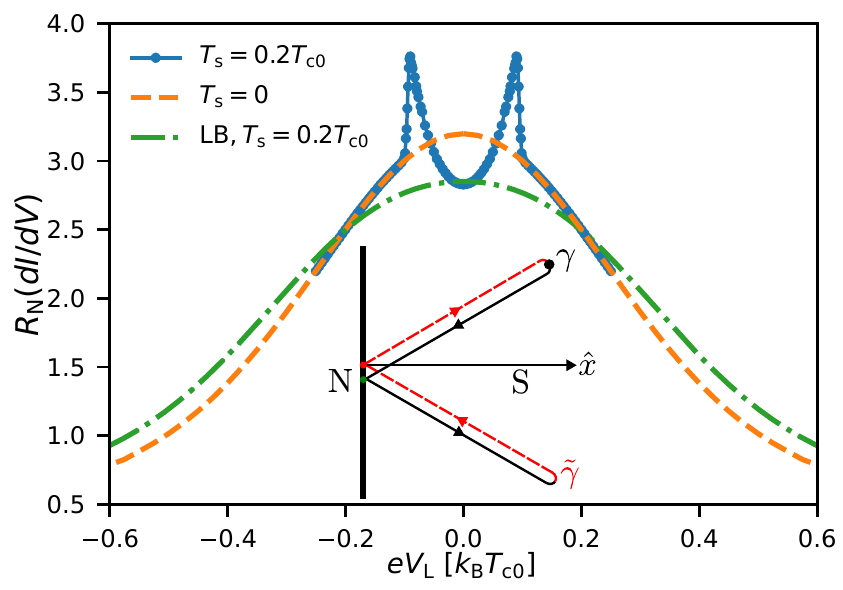}
    \caption{The zero-bias conductance peak (dashed orange line), is a result of the formation of interface Andreev bound states at zero energy. If there is a subdominant $s$-wave component present near the surface, the peak is split due to shifting of the Andreev states to finite energies (solid blue line with circles). The sharpness of the split peak and the rapid fall down back to the pure $d$-wave result at $|eV| = 2 |eV_L| \approx 0.2\kb\tc$ is a result of the suppression and disappearance of the $s$-wave component when a non-equilibrium distribution is enforced under finite voltage bias. In a non-selfconsistent Landauer-B\"uttiker scattering approach, the split is not visible due to thermal broadening (green dash-dotted line), here at $T=0.1\tc$. 
    The inset shows the quasiclassical closed loop of Andreev reflections and normal reflections at the tunnel barrier that leads to the zero-energy Andreev bound state.
    }
    \label{fig:teaser}
\end{figure}

In this paper we return to the problem of tunneling and Andreev reflection spectroscopy of $d$-wave superconducting surfaces including the possibility of a relatively large subdominant $s$-wave component preventing the formation of the phase crystal. We consider the case when the normal metal probe is sufficiently invasive that the non-equilibrium distribution function in the superconductor imposed by the voltage bias has to be computed self-consistently with the order parameter to guarantee current conservation. The voltage dependence of the order parameter, in particular the subdominant component, has a large effect on the spectral properties as well as the transport processes (normal and Andreev reflection amplitudes). Surprisingly, the resulting split of the ZBCP in the presence of the $s$-wave component can be greatly enhanced, as we illustrate in Fig.~\ref{fig:teaser}. The blue dotted line shows the split ZBCP. This non-thermally broadened split is due to the voltage dependence of scattering amplitudes in the presence of the non-equilibrium distribution function in the contact area. As we will show in this paper, the voltage dependence stems from the suppression of the subdominant order under voltage bias. In comparison, in a Landauer-B\"uttiker approach, where the voltage dependence of scattering amplitudes is neglected, the shift of the Andreev states to finite energy is not visible in the conductance due to thermal broadening, see the dash-dotted green line in Fig.~\ref{fig:teaser}. This physics becomes relevant for geometries of the normal-metal-insulator-superconductor (NIS) contact where the traditional point-contact assumption can not be made due to the largeness of the contact, with a diameter larger than the small superconducting coherence length in the high-temperature superconductor, and relatively high transparency of the tunnel barrier. In this case, the non-equilibrium distribution has to be computed self-consistently to guarantee current conservation. The superconducting order parameters as well as impurity self-energies then render the scattering amplitudes voltage-dependent, thus influencing the conductance.

The paper is organized as follows. In section \ref{sec:model} we outline the model assumptions we make for calculating the stationary conductance-voltage characteristics including current-conservation. The details of the quasiclassical theory that we use have been summarized in Appendix \ref{sec:Appendix}. In the following section \ref{sec:results} we go, step by step, through results for the current-conserving theory for first the pure $d_{x^2-y^2}$ case for two orientations of the order parameter relative to the tunnel barrier normal. In section \ref{sec:alpha0} for zero misorientation (the [100] orientation) there are no zero-energy Andreev bound states. In this case we discuss the effect of charge imbalance in a $d$-wave superconductor as compared to the more well studied conventional $s$-wave case \cite{clarke_experimental_1972,tinkham_theory_1972,artemenko_electric_1979,arutyunov_relaxation_2018,Seja2021Sep}. In section \ref{sec:alpha45} we turn to the case of 45$^\circ$ misorientation (the [110] orientation), and discuss the influence of the zero-energy Andreev bound states on charge imbalance and the influence of non-equilibrium on the ZBCP. Lastly, in section \ref{sec:d_plus_is} we include the subdominant $s$-wave order parameter in the energetically favorable time-reversal symmetry-breaking combination $d_{x^2-y^2}+is$. We then study in more detail the conductance-voltage characteristics shown in Fig.~\ref{fig:teaser}. The paper is summarized in section \ref{sec:summary}.

\section{Model and methods}\label{sec:model}

\begin{figure}[t]
    \centering
    \includegraphics[width=\columnwidth]{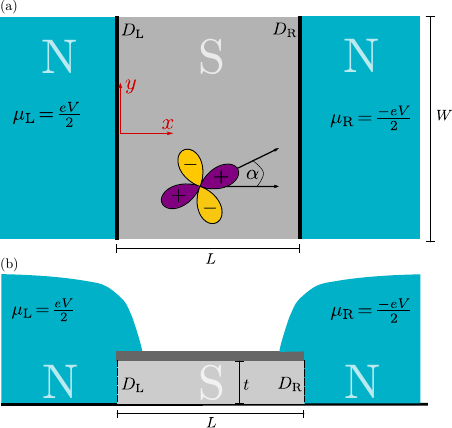}
    \caption{Principle setup of the model. A $d$-wave superconducting film of thickness $t$ (grey) is connected to two reservoirs at $\xco=0$ and $\xco=L$ via barriers with transparencies $D_\mathrm{L}$ and $D_\mathrm{R}$ (black-hatched surfaces). Translational invariance is assumed in the $y$-direction and homogeneity is assumed in the perpendicular $z$-direction. The left reservoir is at potential $\mu_\mathrm{L} = eV/2$ while the right reservoir is at $\mu_\mathrm{R} = -eV/2$. Both are at temperature $T$. The angle $\alpha$ specifies the crystal axis misalignment with respect to the main transport axis, $\hat x$. We will limit the discussion to a symmetric system with $D_\mathrm{L}=D_\mathrm{R}=D$. The sideview corresponds to the experimental techniques \cite{Baghdadi2015Jul,Baghdadi2017} to contact the film in the $ab$-plane direction with a protecting capping layer on top.}
    \label{fig:model_voltbiased_wire}
\end{figure}

A sketch of the setup that we are considering is given in Fig.~\ref{fig:model_voltbiased_wire}. A $d$-wave superconductor (S) of length $L$ and width $W\gg L$ is connected through tunnel barriers (I) to two normal metal reservoirs (N) held at voltages $\pm eV/2$, both at a base temperature $T$. We assume a thin film superconductor, thin in the perpendicular $z$-axis direction, with thickness $t$ and transport from left to right along the $\xco$-axis, which is parallel to the normals of the NIS interfaces. We consider injection of carriers into the superconducting film edges in the $ab$-plane direction, corresponding to experiments where such high-transparency contacts between YB$_2$Cu$_3$O$_{7-\delta}$ and Au have been fabricated \cite{Baghdadi2015Jul,Baghdadi2017}. The film thickness $t$ is assumed small compared with the penetration depth $\lambda_\mathrm{c}$ in the $c$-axis direction (along film normal), and contacts are assumed homogeneous. The current density can then be assumed to be the same in all layers of the film and we may consider a single two-dimensional layer in the following. We consider relatively low temperatures, and assume that the in-plane penetration depth $\lambda$ is large compared with the coherence length $\xi_0=\hbar v_\mathrm{F}/2\pi\kb\tc$, where $\hbar$ is the reduced Planck constant, $\kb$ is the Boltzmann constant, $v_F$ is the normal state Fermi velocity, and $\tc$ is the critical temperature. In this case for a superconductor of width $W$ fullfilling $\xi_0\ll W\ll\lambda$, screening can be neglected and the charge current can be considered to flow homogeneously as function of the transverse $y$ coordinate. The assumption of translational invariance breaks down near the lower ($y=0$) and upper edges ($y=W$), but we assume that the contributions to the total current from these edges are small compared to the large translationally invariant flow in the interior. In fact, the restriction on system size is weaker since for the thin film of thickness $t<\lambda$ the pearl length $\lambda_\mathrm{p}=\lambda^2/t\gg\lambda$ guarantees homogeneous flow in the transverse $y$-direction as long as $\xi_0\ll W\ll\lambda_\mathrm{p}$.

We utilize the non-equilibrium quasiclassical theory of superconductivity. An overview over the relevant equations is given in Appendix \ref{sec:Appendix}, see also Refs.~\cite{Seja2021Sep,Seja2022Mar} for more details. In short, the Eilenberger-Larkin-Ovchinnikov equations \cite{Eilenberger1968Apr,Larkin1969, Eliashberg1971} for the Keldysh, retarded, and advanced quasiclassical Green's functions, Eq.~\eqref{eq:transportequation}, are solved self-consistently with the superconducting order parameter $\Delta$, impurity self-energies for arbitrary mean free path $\ell$, and the local electrochemical potential $\phi(\xco)$.
The self-consistent treatment guarantees that the charge current $j$ is conserved from source to drain contacts, i.e., it is independent of $\xco$.

The singlet $d$-wave order parameter is written as
\begin{equation}
\Delta(\pF,\RR)=\Delta_\mathrm{d}(\RR)\eta_\mathrm{d}(\pF),
\label{eq:OrderParameterEquation}
\end{equation}
where the orbital basis function $\eta_\mathrm{d}(\pF)=\sqrt{2}\cos\left[2\left(\varphi_\mathrm{F}-\alpha\right)\right]$ depends on $\varphi_\mathrm{F}$, which is the angle between the Fermi momentum $\pF$ and the $\xco$-axis, and the angle $\alpha$ giving the orientation of the $d$-wave clover with respect to the $\xco$-axis, see Fig.~\ref{fig:model_voltbiased_wire}. The amplitude is in non-equilibrium a spatially dependent complex quantity, $\Delta_\mathrm{d}(\RR)=|\Delta_\mathrm{d}(\RR)|\exp[i\chi(\RR)]$. Through a local gauge transformation, one may transform the order parameter to be real, and thereby obtain the superfluid momentum $\ps=\tfrac{\hbar}{2}\nabla\chi$ which signals the presence of superflow.

The superconductor is connected to the normal-metal reservoirs via insulating barriers, which are assumed to be symmetric, $D_\mathrm{L} = D_\mathrm{R} =D$.
In this paper, we include a tunnel cone such that the interface transparency $D$ depends on the momentum angle $\varphi_\mathrm{F}$. Explicitly, we use  
$D(\varphi_\mathrm{F})=D_0(e^{-\beta \sin^2 \varphi_\mathrm{F}} - e^{-\beta})/(1 - e^{-\beta})$.
Here, $D_0$ is the transparency for perpendicular incidence on the surface, and the parameter $\beta$ determines the narrowness of the tunnel cone. Throughout this paper, we use $\beta = 1$ corresponding to a wide tunnel cone. 
\begin{figure*}[th!]
    \centering
    \includegraphics[width=\textwidth]{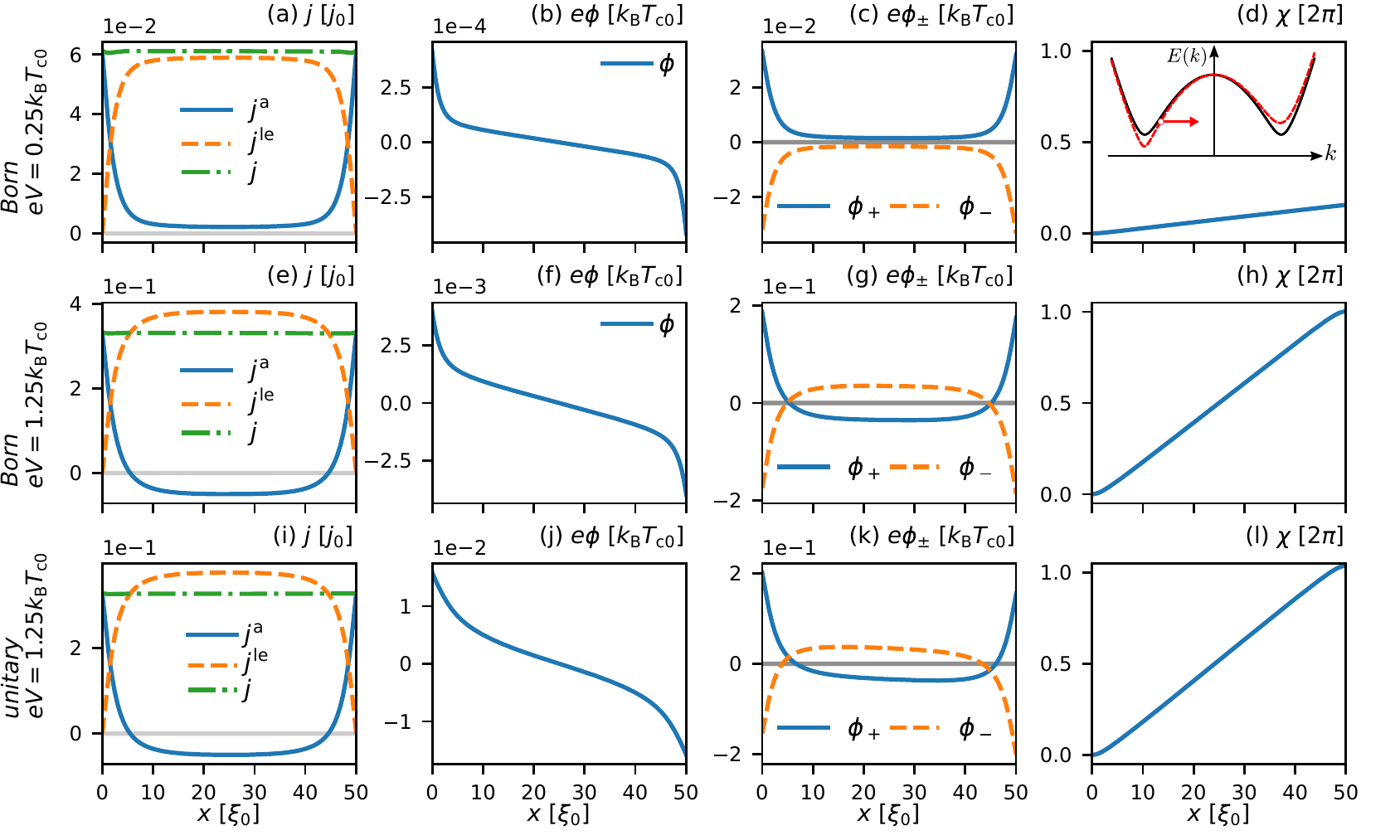}
    \caption{Spatial dependencies of main physical quantities and observables for high-transparency interfaces $D_0=1$ with wide tunnel cones ($\beta = 1$), orientation $\alpha = 0$, mean free path $\ell=100\xi_0$, and $T=0.1 T_\mathrm{c}$. The first and second row show results for the Born limit at applied bias $eV=0.25\kb\tc$ (top row) and $eV=1.25\kb\tc$ (bottom row). The third row shows results at the larger bias for impurities scattering in the unitary limit. 
    First column: anomalous and local-equilibrium current densities, $j^\mathrm{a}(\xco)$ and $j^\mathrm{le}(\xco)$, see Eq.~\eqref{eq:CurrentSplitting}. They add up to a conserved total current $j(\xco)=\mathrm{constant}$.
    Second column: local quasiparticle potential $\phi(\xco)$.
    Third column: left-mover and right-mover electrochemical potentials, $\phi_+(\xco)$ and $\phi_-(\xco)$.
    Fourth column: superconducting phase drop $\chi(\xco)$.
    The inset in (d) shows a sketch of the superconducting quasiparticle dispersion (black solid line), with the Doppler shifting effect of superflow as a dashed red line. The right-pointing arrow indicates a hole-like quasiparticle state with positive group velocity.
    }
    \label{fig:overviewBornUnitary}
\end{figure*}

The impurity self-energies for a dilute concentration $n_\mathrm{imp}$ of impurities is computed using a self-consistent $t$-matrix approximation, including multiple scattering off individual impurities but neglecting crossing diagrams. As seen in Eqs.~\eqref{eq:tRDefinition}-\eqref{eq:tKDefinition}, the model contains two parameters, the impurity concentration $\Gamma_u=n_\mathrm{imp}/\pi\NF$ and the scattering phase shift, $\delta_0=\arctan(\pi \NF u_0)$, assuming an isotropic impurity potential $u_0$. 

We consider here two limits for the potential strength. In the Born limit, $u_0$ is small such that only the first diagram in the $t$-matrix series is kept. In the second unitary limit, $u_0$ is considered very large and the t-matrix equation is summed to infinite order for multiple scattering off a single impurity. In both cases a single parameter $\Gamma=\Gamma_u
\sin^2\delta_0$, related to the normal state mean free path $\ell=\hbar v_\mathrm{F}/2\Gamma=(\pi/\Gamma)\xi_0$, characterizes the impurities. In the Born limit we assume $u_0$ small, but $n_\mathrm{imp}$ large, keeping $\Gamma$ constant, while in the unitary limit $u_0$ is large such that $\delta_0\rightarrow\pi/2$. This homogeneous scattering model has been widely studied in unconventional, in particular $d$-wave, superconductors \cite{AGD:Book,arf88,xu95,graf96,Poenicke1999,LofwanderProximity2004,Lofwander2004Jul,Seja2021Sep,Seja2022Mar}. In the bulk superconductor, the Born limit impurities simply broaden the density of states, while the unitary limit impurities introduce a sizeable impurity band of resonance states around the Fermi energy. This low-energy band of quasiparticles influences transport. At the surface, when there are Andreev bound states at zero energy for misorientation angle $\alpha\neq 0$, the Born limit impurities heavily broadens them, while the unitary limit impurities are less effective in increasing their lifetime.

The $\phi(\xco)$-potential, see Eq.~\eqref{eq:phi_hmatrix}, quantifies charge imbalance, i.e., the difference in chemical potential locally at $\xco$ between the condensate and the quasiparticle states. This potential is determined through the assumption of local charge neutrality \cite{Eschrig1999Oct}: excess charge due to injection of quasiparticles is compensated by a local depletion of the condensate. This also means that we neglect any charging effects: the charging energy $U$ of the central region is small compared with the energy scale $\hbar/\tau_\mathrm{d}$ ($\tau_\mathrm{d}$ dwell time) due to escape to the leads. The charge imbalance is induced at the interfaces when quasiparticles are injected from the normal metals and decays into the interior of the superconductor. The processes determining the decay away from the interfaces are Andreev reflection and impurity scattering, which is particularly important for the $d$-wave order parameter with gap nodes. We find that due to the gap nodes, charge imbalance does not decay to zero for any voltage $|V|>0$. This is the case both for ballistic ($\ell>L$) and diffusive devices ($\ell<L$). Experimentally, for sufficiently long devices, inelastic processes will relax charge imbalance. In our treatment, we assume that the inelastic mean free path is large both compared to elastic mean free path $\ell_\mathrm{in}\gg\ell$ and system size $\ell_\mathrm{in}\gg L$. Thus, the dwell time of non-equilibrium quasiparticles in the superconductor is considered small compared with inelastic relaxation times \cite{Seja2021Sep}.

\section{Results}\label{sec:results}

\subsection{Orientation $\alpha=0$}\label{sec:alpha0}

\begin{figure}
    \centering
    \includegraphics[width=\columnwidth]{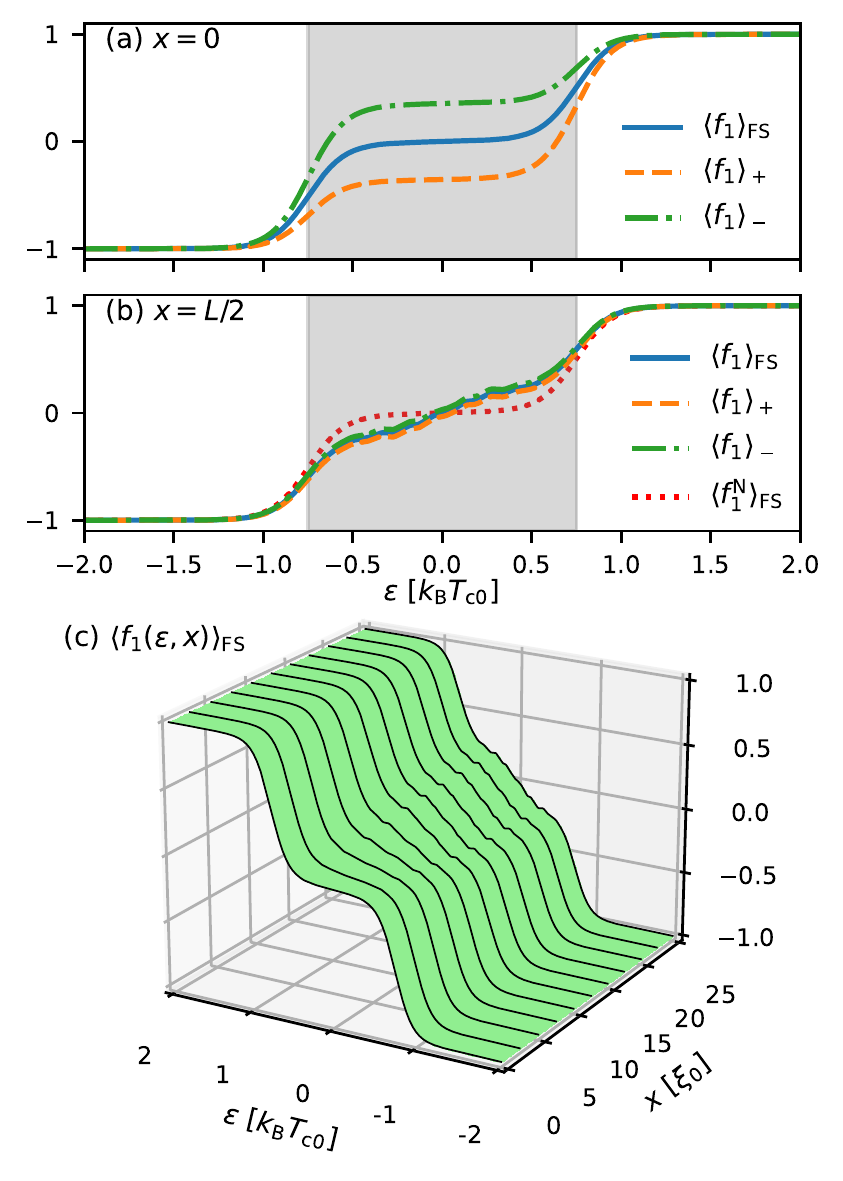}
    \caption{The energy-like mode $f_1$, see Eq.~\eqref{eq:h-splitting}, averaged over the full Fermi surface, $\FS{f_1(\pF,x,\varepsilon)}$, and over half the Fermi-surface with positive momenta $\FSplus{f_1(\pF,x,\varepsilon)}$ and negative momenta $\FSminus{f_1(\pF,x,\varepsilon)}$ with respect to the $\xco$-axis, see also Eq.~\eqref{eq:PartialFermiAverage}. In (a) at the N-S interface ($\xco=0$), in (b) at the center of the superconductor ($\xco=L/2$), the grey shaded area in both figures indicates the voltage window of width $eV = 1.5 \kb \tc$. In (c), we show the energy-resolved spatial evolution of $\FS{f_1(\varepsilon, \xx)}$ from the surface to the center of the system. In all cases, we have  $D_0=1$, $\beta = 1$, $\alpha=0$, Born limit impurities with $\ell=100\xi_0$, and $T=0.1T_c$.}
    \label{fig:distributions}
\end{figure}

In Fig.~\ref{fig:overviewBornUnitary} we show an overview of the spatial dependencies of all relevant physical quantities for the orientation $\alpha=0$ and a large mean free path $\ell=100\xi_0$ with impurity scattering in the Born limit for one low and one high voltage, and for the unitary limit at high voltage only. For high transparency of the interface barriers, $D_0=1$ and wide tunnel cones ($\beta=1$), transport in the contact regions are dominated by Andreev reflection at low voltage. This leads to a conversion of quasiparticle flow [anomalous component $j^\mathrm{a}(\xco)$] at the interfaces to a dominating superflow [local equilibrium component $j^\mathrm{le}(\xco)$] far from the contacts, see Fig.~\ref{fig:overviewBornUnitary}(a). The main qualitative difference from the conventional $s$-wave superconducting case \cite{Seja2021Sep} is the presence of the $d$-wave gap nodes and the more strong pair-breaking effect of elastic impurity scattering. This leads to a charge imbalance throughout the system for all voltages $|V|>0$, characterized by a potential $\phi(\xco)$, see Fig.\ref{fig:overviewBornUnitary}(b), as well as a residual quasiparticle flow [anomalous component $j^\mathrm{a}(\xco)$] in the interior of the superconductor, see Fig.~\ref{fig:overviewBornUnitary}(a). It was shown for conventional but anisotropic $s$-wave superconductors that charge imbalance may be relaxed by elastic impurity scattering \cite{tinkham_theory_1972}. For the $d$-wave case, we find that after an initial drop of $\phi(\xco)$ through Andreev processes in the contact region, the gap nodes (absence of a true energy gap) and pair-breaking elastic impurity scattering result in a residual charge imbalance extending into the interior of the superconductor for all system sizes $L\lesssim 200\xi_0$ that we have considered. Note however, that in our setup the potential profile necessarily is antisymmetric with respect to the center of the system and $\phi(\xco)$ is forced to pass through zero at the center.

At higher bias, the superconducting phase gradient leads to substantial Doppler shifts of the continuum states. As illustrated in the inset of Fig.~\ref{fig:overviewBornUnitary}(d), right-moving hole-like quasiparticle states are shifted down into the bias window, while right-moving electron-like quasiparticle states are shifted up. As a consequence, an electron- to hole-like quasiparticle transmission process $T_\mathrm{he}$ becomes available in the interface region \cite{sanchez-canizares_self-consistent_1997,Seja2021Sep}. This effect leads to hole-like quasiparticle transport from left to right in the figure and a negative quasiparticle current $j^\mathrm{a}(\xco)<0$ is induced in the interior of the superconductor, see Fig.~\ref{fig:overviewBornUnitary}(e). An associated sign change in right- and left-mover quasipotentials $\phi_{+}(z)$ and $\phi_{-}(z)$, see Eq.~\eqref{eq:PhiLeftRight}, also appears, see Fig.~\ref{fig:overviewBornUnitary}(g). The condition of current conservation then forces the condensate to compensate for this through larger phase gradients and a resulting larger local-equilibrium current, $j^\mathrm{le}(\xco)>j$, see Fig.~\ref{fig:overviewBornUnitary}(e). Eventually, at higher voltage, but at a voltage below the maximum of the $d$-wave superconducting gap, superconductivity in the mesoscale device breaks down. The breakdown is due to the combined effect of Doppler shifting superflow and the highly non-equilibrium form of the distribution function $f_1(\pF,\RR,\varepsilon)$ that enters the gap equation Eq.~(\ref{eq:DeltaSelfConsistencyEq_hmatrix}), see also Fig.~7 of Ref.~\cite{Seja2021Sep} and the discussion in Ref.~\cite{Keizer2006Apr}.
As compared with the $s$-wave superconducting case, the break down is at a lower voltage, due to the $d$-wave gap nodes. We also note that the breakdown appears at a lower current compared with the usual critical current due to pure superflow \cite{Bardeen1962}.

In Fig.~\ref{fig:distributions} we show the distribution function $f_1(\pF,x,\varepsilon)$ for a voltage $eV=1.5\kb\tc$ below the maximum of the $d$-wave gap. As defined in Eq.~\eqref{eq:PartialFermiAverage}, the Fermi surface average has been divided into positive and negative projections of the Fermi momenta on the transport $\xco$-axis, $\left\langle f_1(\pF,x,\varepsilon) \right\rangle_{\pm}$, reflecting injection of electrons from the left and right reservoirs. The distribution is modified from its equilibrium form $f_1^\mathrm{eq}(\varepsilon)=\tanh(\varepsilon/2T)$ into a characteristic two-step shape at the contact, where the step width is given by the local potential, here $eV/2$. In the interior of the superconductor, the distribution remains highly non-equilibrium and cannot be characterized by a local effective temperature, since it does not have the shape of the $\tanh$-function. Instead, it can be understood in simplified terms as the result of a superposition of a superconducting and a normal component due to the $d$-wave gap nodes. The normal component would be a constant through the superconductor, given by 
$f_1^\mathrm{N}(\varepsilon)=\tfrac{1}{2}\left(
 \tanh[(\varepsilon-eV/2)/(2T)]
+\tanh[(\varepsilon+eV/2)/2T)] \right)$,
[red dotted line in Fig.~\ref{fig:distributions}(b)] while the superconducting component has a spatial dependence such that it relaxes back to the equilibrium form $\tanh(\varepsilon/2T)$ away from the contacts once Andreev processes are complete. The self-consistently computed distribution in Fig.~\ref{fig:distributions} is a result of an interplay between these two components. The finite quasiparticle flow in the interior of the superconductor, the finite $j^\mathrm{a}(\xco)$ in Fig.~\ref{fig:overviewBornUnitary}(a), is reflected in the difference between the distributions with positive and negative projections.
The superconductivity is weakened when the distribution function $f_1(\pF,x,\varepsilon)$ is reduced substantially in a window of subgap energies, as marked by the grey shaded area in Fig.~\ref{fig:distributions}. Superconductivity breaks down when the window of reduced distribution reach the superconducting coherence peaks.

Let us next consider strong impurity scattering. In the case of unitary-limit scattering, there is an impurity band formed around the Fermi energy of width $\sim\sqrt{\pi\Delta_\mathrm{d}\Gamma/2}$ \cite{lee93,graf96}. This enhances charge imbalance as compared with the Born limit for the same mean free path, since the superconductor is more normal metal like. The residual potential $\phi(\xco)$ in the interior superconductor is therefore enhanced, see Fig.~\ref{fig:overviewBornUnitary}(j).

The moving condensate of the superconductor is associated with a superconducting phase gradient.
An alternative view is to use a gauge where the order parameter is real, in which case there is a spatially dependent superfluid momentum $p_\mathrm{s}(\xco)=\tfrac{\hbar}{2}\partial_x\chi(\xco)$.
As a result, a phase difference $\Delta\chi=\chi(L)-\chi(0)$ is established across the structure, see the right-most column of Fig.~\ref{fig:overviewBornUnitary}. For larger voltages, this phase difference can grow large and surpass $2\pi$. Since the superconductor is rather long ($L=50\xi_0$ in Fig.~\ref{fig:overviewBornUnitary}), the supercurrent is still below the bulk depairing current, $p_\mathrm{s}<p_\mathrm{sc}$, where $p_\mathrm{sc}$ is the critical superfluid momentum where superconductivity breaks down due to superflow \cite{Bardeen1962}.

\begin{figure}[t]
    \centering
    \includegraphics[width=\columnwidth]{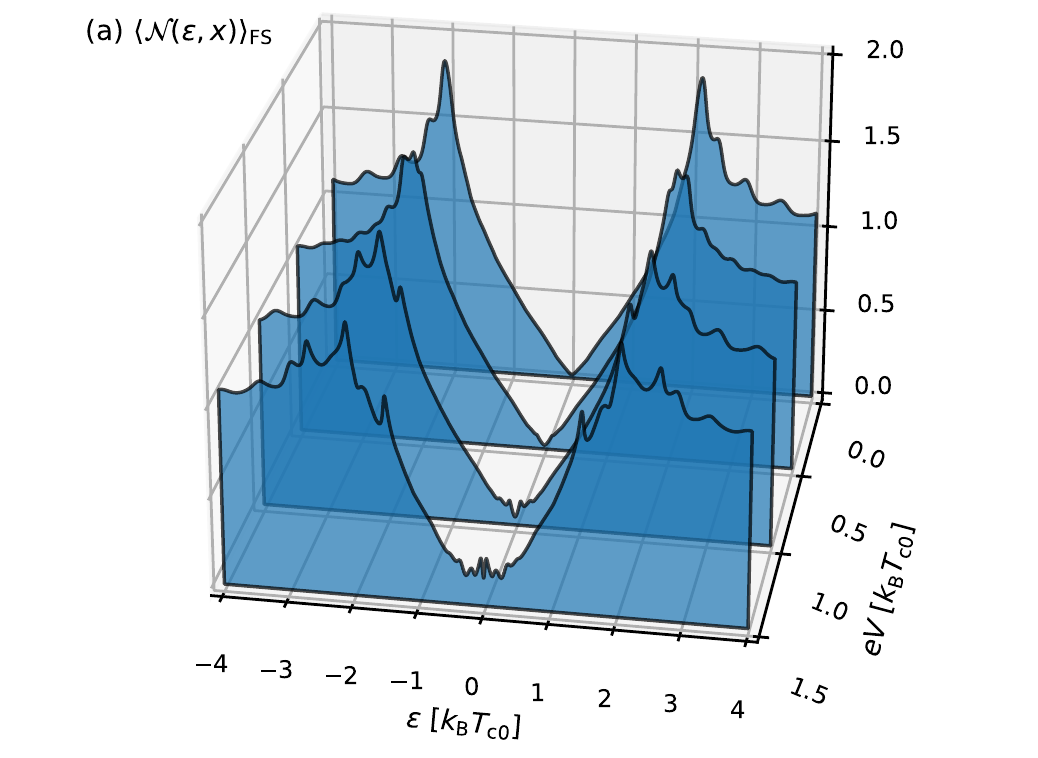}
    \caption{Fermi-surfaced averaged density of states $\FS{\mathcal{N}(\varepsilon, x) }$, see Eq.~\eqref{eq:DensityOfStates_averaged}, at the center of the superconductor for different voltages for the same parameters as in Fig.~\ref{fig:distributions}.}
    \label{fig:dos}
\end{figure}

To further quantify the disequilibrium throughout the superconductor we consider the local density of states (LDOS) under voltage bias, see Fig.~\ref{fig:dos}. Already in equilibrium, the $d$-wave DOS is modified in a wide region near the contact due the inverse proximity effect \cite{LofwanderProximity2004}. Under voltage bias, the main changes in the LDOS are due to the Doppler shifts from the moving condensate. In simplified terms, for a clean system, the superconducting coherence peak at the maximum of the $d$-wave gap is expected to split into two peaks, for quasiparticle states co-moving and counter-moving with the condensate flow \cite{xu95}. The picture changes due to impurity scattering, which mixes the co- and counter-moving quasiparticle states present in a superclean system. For more dirty systems, for instance $\ell\sim 10\xi_0$ (not shown in the figure), impurity scattering broadens the substructure into a wide peak around the gap energy. At low energies there is an enhanced density of states within a window $|\varepsilon|\sim v_\mathrm{F}p_\mathrm{s}(\xco)$, which is also due to the Doppler shifts.

\subsection{Orientation $\alpha=\pi/4$}\label{sec:alpha45}

\begin{figure*}[t]
    \centering
    \includegraphics[width=\textwidth]{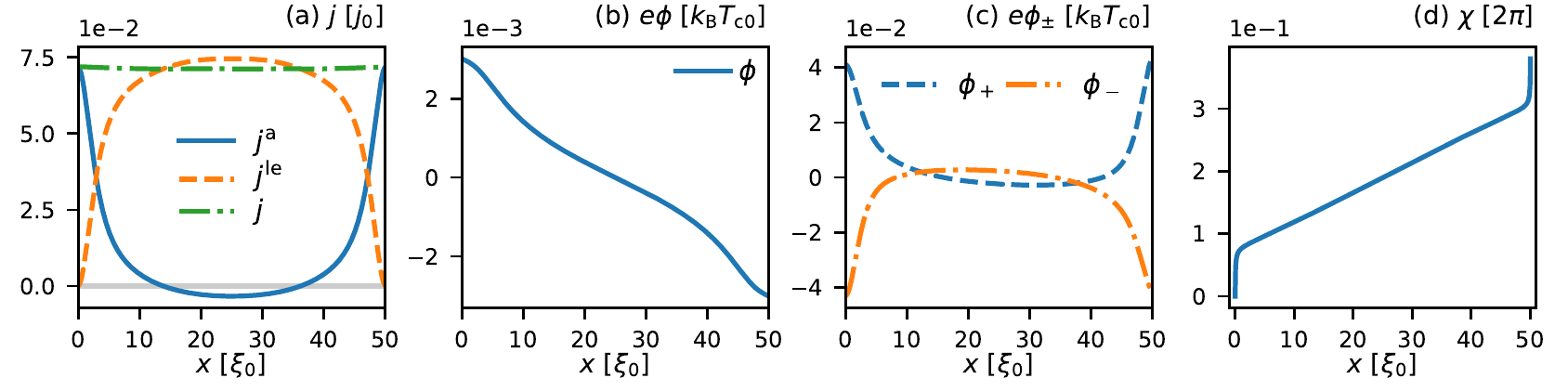}
    \caption{Spatial dependencies of main physical quantities for the orientation $\alpha=\pi/4$ where zero-energy surface states are formed.
    (a) Anomalous and local equilibrium current densities, $j^\mathrm{a}(\xco)$ and $j^\mathrm{le}(\xco)$. (b) Local quasiparticle potential $\phi(\xco)$. (c) Left-mover and right-mover potentials, $\phi_+(\xco)$ and $\phi_-(\xco)$. (d) Superconducting phase drop $\chi(\xco)$.
    The interface transparency is $D_0 = 0.2$ with a wide tunnel cone ($\beta = 1$), scattering is in the unitary limit with a mean free path $\ell=100\xi_0$, the applied bias $eV = 1\kb \tc$, and $T=0.2 T_\mathrm{c}$.
    }
    \label{fig:overviewAlpha45}
\end{figure*}
\begin{figure}
    \centering
    \includegraphics[width=\columnwidth]{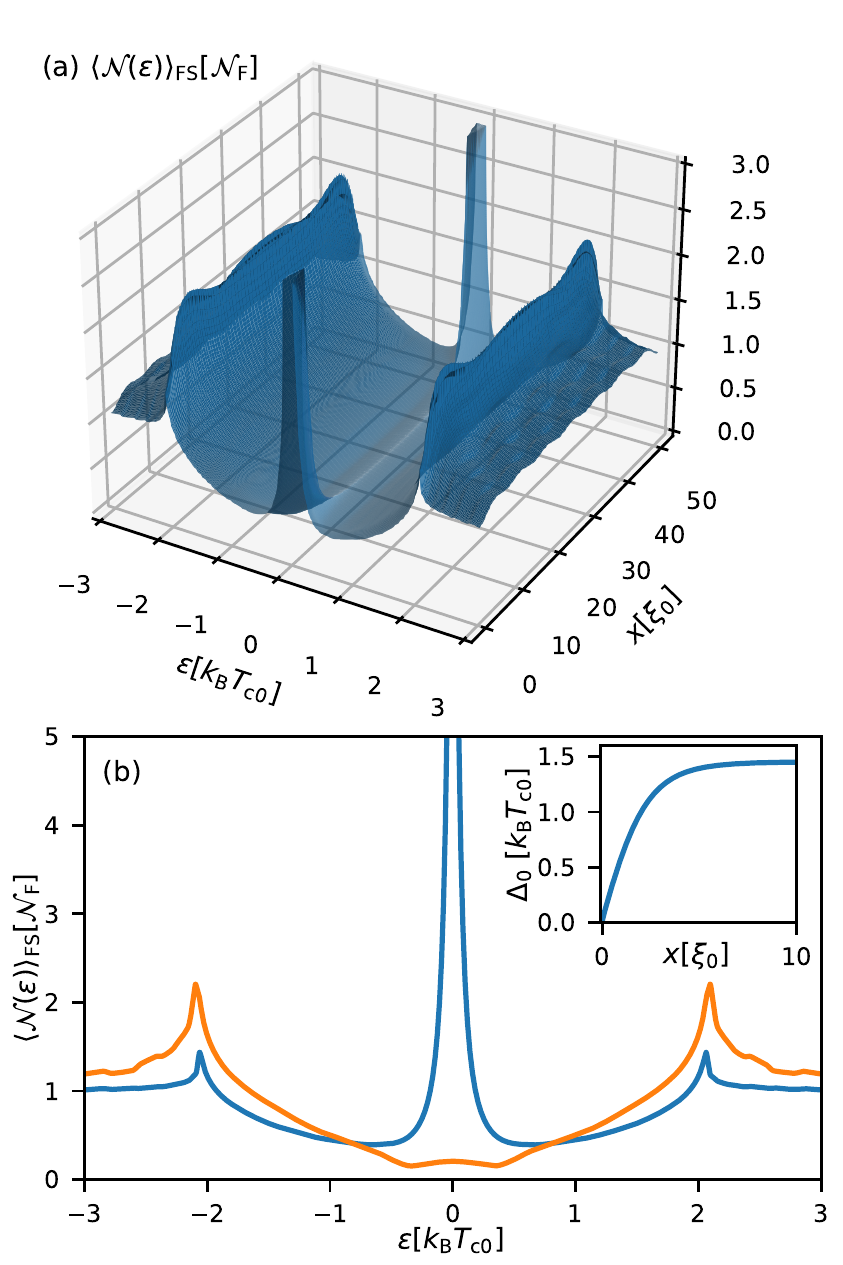}
    \caption{(a) The density of states as functions of energy and spatial coordinate for the same parameters as in Fig.~\ref{fig:overviewAlpha45} but in equilibrium. (b) Cut of the density of states at the interface, $\xco=0$ (blue line), and in the center of the superconductor, $\xco=L/2$ (orange line). The inset shows the suppression of the order parameter close to the surface. 
    }
    \label{fig:dos_boundstate}
\end{figure}

At a surface of a $d$-wave superconductor, with the surface normal misaligned with the crystal main axes ($\alpha \neq 0$), Andreev bound states at zero energy, the so-called midgap states, are formed \cite{Hu1994}. For the case of $\alpha = \pi/4$, such zero-energy states exist for all trajectory angles $\varphi_\mathrm{F}$. For an interface to a normal metal, the zero-energy peak in the interface LDOS acquires a width determined by the interface transparency, $\sim\FS{\Delta(\varphi_\mathrm{F})D(\varphi_\mathrm{F})}$. In the presence of impurity scattering, the LDOS peak is broadened further \cite{Poenicke1999}. 
A clear signature of the zero-energy states is a peak in the differential conductance around zero voltage in a point-contact geometry, as also found experimentally \cite{kashiwaya_tunnelling_2000}. Here, we go beyond the point-contact assumption and investigate the influence of zero-energy states on charge imbalance, as well as the influence of a non-equilibrium distribution on the zero-bias conductance peak.

In Fig.~\ref{fig:overviewAlpha45} we show the spatial dependencies of all relevant physical quantities for the case of unitary impurity scattering with $\ell=100\xi_0$, low-transparency interfaces $D_0 = 0.2$, and a voltage bias of $eV=1.0\kb\tc$. In this case, there are both zero-energy Andreev bound states at the interfaces to the normal metals and an impurity band around zero energy in the interior of the superconductor. The combination enhances charge imbalance and the $\phi$-potential, see Fig.~\ref{fig:overviewAlpha45}(b). The formation of the Andreev bound states is associated with a suppression of the order parameter near the interfaces, see inset in Fig.~\ref{fig:dos_boundstate}(b). This weakening of superconductivity forces the condensate to produce a large phase gradient to set up the necessary superflow, see Fig.~\ref{fig:overviewAlpha45}(d). The resulting superfluid momentum $\ps(\xco)=p_\mathrm{s}(\xco){\hat x}$  with $p_\mathrm{s}(\xco)=\tfrac{\hbar}{2}\partial_x\chi(\xco)$ grows large near the contacts.

The effect of superflow for $\alpha=0$ considered above is best understood in terms of Doppler shifts $\varepsilon\rightarrow\varepsilon-\vF\cdot\ps$ of continuum quasiparticles propagating along trajectories characterized by a distinct $\vF$. For surface Andreev states this is not the case when the superflow is along the surface normal $\ps\parallel{\hat x}$ as in our case, since the Andreev bound states are a superposition of partial waves with both positive and negative projections of the Fermi momentum (and Fermi velocity) with respect to the $\xco$-axis, see inset in Fig.~\ref{fig:teaser}.
This leads to different signs of the shifts $\vF\cdot\ps$ for the partial waves with opposite momentum projections on the $\xco$-axis, and there is no resulting Doppler shift. Instead, the finite superflow only changes the spectral weight of these states. 
For an illustration of this physics, we assume a clean $d$-wave superconductor with a perfectly reflective interface at $\xco=0$ and neglect the suppression of the order parameter close to the surface. For a Fermi velocity with angle $\varphi_\mathrm{F} \in (-\pi/2, \pi/2)$, specular scattering at the interface connects the two velocities
\begin{align}
\mathbf{v}_\mathrm{F}^\mathrm{out}\!=\! 
\begin{pmatrix}
v_\mathrm{F}^x\\
v_\mathrm{F}^y
\end{pmatrix}
\equiv 
v_\mathrm{F}\begin{pmatrix}
\cos \varphi_\mathrm{F}\\
\sin \varphi_\mathrm{F}
\end{pmatrix}
,
\mathbf{v}_\mathrm{F}^\mathrm{in} = 
v_\mathrm{F}
\begin{pmatrix}
-\cos \varphi_\mathrm{F}\\
\sin \varphi_\mathrm{F}\\
\end{pmatrix}.
\label{eq:vF_x_definition}
\end{align}
The order parameter with $\alpha=\pi/4$ has opposite signs along those two direciton, see Eq.~\eqref{eq:OrderParameterEquation}.
The resulting surface retarded Green's function reads
\begin{align}
\frac{g^\mathrm{R}(z,\varphi_\mathrm{F})}{-i\pi} =
\frac{ (\vfin - \vfout) \cdot \ps + i \left[\Omega(\vfout) + \Omega(\vfin) \right] 
}
{2z - (\vfin + \vfout) \cdot \ps + i \left[\Omega(\vfout) - \Omega(\vfin)  \right]},
\label{eq:DensityOfStates_ABS}
\end{align}
where
\begin{align}
\Omega(\vF) \equiv \sqrt{ \Delta^2(\pF) - (z - \vF \cdot \ps )^2 }, 
\end{align}
and $z=\varepsilon+i0^+$ is the energy with an infinitesimal positive imaginary part.
In the absence of $\ps$, it is straightforward to show that Eq.~\eqref{eq:DensityOfStates_ABS} has only one first-order pole at $z=0$ with a residue of $\pi |\Delta(\varphi_\mathrm{F})|$, which is the spectral weight of the Andreev bound state for this trajectory angle. For a general $\ps=(p_\mathrm{s}^x,p_\mathrm{s}^y)^\mathrm{T}$, an identical analysis shows that as long as $|p_\mathrm{s}^x| < |\Delta|$, the pole is shifted to $z = v_\mathrm{F}^y p_\mathrm{s}^y$, while the residue is reduced to 
\begin{align}
\pi\sqrt{\Delta^2(\pF) - (v_\mathrm{F}^x p_\mathrm{s}^x )^2 }.
\end{align}
Thus, we see that the two components of $\ps$ have very different influence on the Andreev bound state.
The component normal to the interface, $p_\mathrm{s}^x$, reduces the spectral weight of the Andreev state. This is compensated for by an enhanced spectral weight of the continuum. In contrast, the component parallel to the surface, $p_\mathrm{s}^y$, shifts the state in energy according to $\varepsilon \rightarrow \varepsilon + v_\mathrm{F}^y p_\mathrm{s}^y$.
As a result, the zero-energy states shift to positive (negative) energy for trajectories with positive (negative) projection on the $y$-axis, and the single peak in the density of states splits into two separate peaks. This is indeed the case in the presence of an external magnetic field along the $c$-axis direction perpendicular to the superconducting film, where the induced screening currents Doppler shift and split the ZBCP as found experimentally \cite{Covington1997Jul}. In our case, though, the voltage bias only results in superflow along the $\xco$-axis and the Andreev states stay at zero energy.

\begin{figure}
    \centering
    \includegraphics[width=\columnwidth]{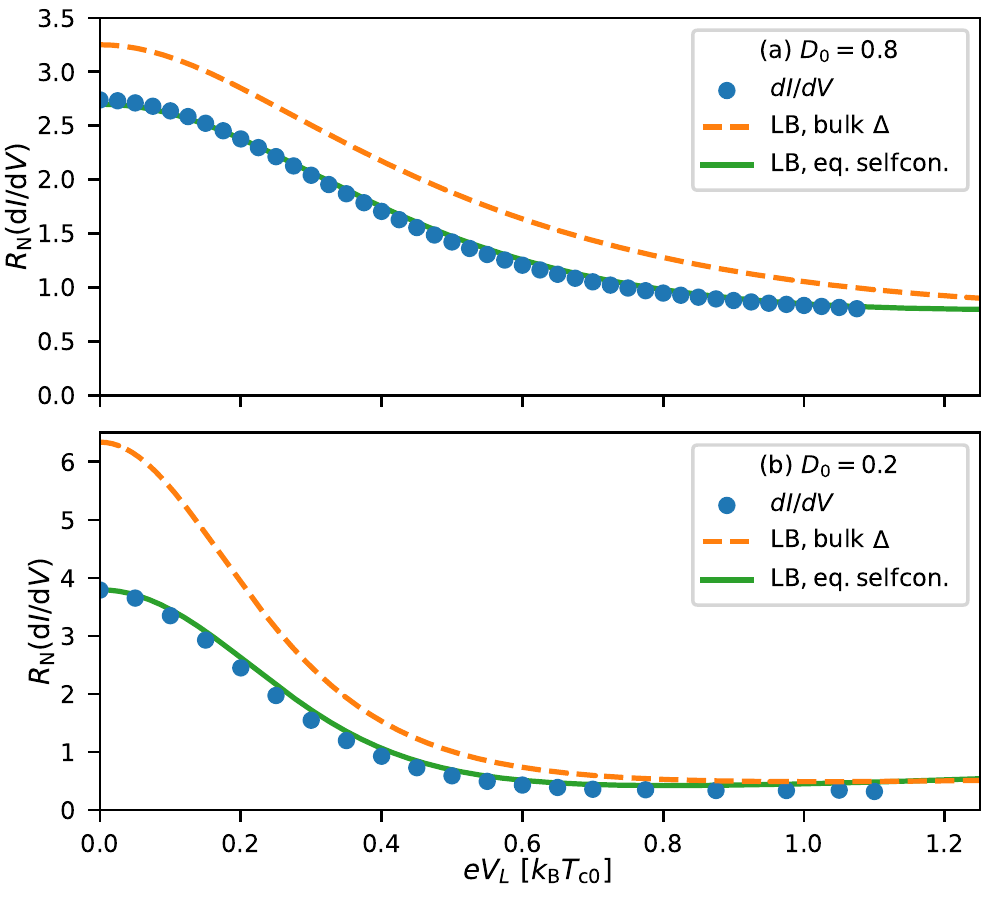}
    \caption{Differential conductance $dI/dV$, normalized to the normal-state interface resistance, in comparison to a Landauer-B\"uttiker scattering approach in the spirit of Ref.~\onlinecite{Tanaka1995Apr}. The interface transparencies are (a) $D_0 = 0.8$, and (b) $D_0 = 0.2$. In both cases $\alpha=\pi/4$, $T=0.1\tc$, and $\ell=100\xi_0$ in the Born limit.
    Note that all horizontal axes show $eV_\mathrm{L} = eV/2$ for our model system, corresponding to the applied bias at the (single) interface in the scattering approach.
    }
    \label{fig:conductance_comparison}
\end{figure}

In Fig.~\ref{fig:conductance_comparison} we show results for the conductance as function of applied voltage for the orientation $\alpha=\pi/4$ and two different transparencies of the interface barriers. The conductance is computed by numerical differentiation of the current defined in Eq.~(\ref{eq:ChargeCurrentDefinition}), and is independent of the coordinate $\xco$. For an analysis of the current and conductance it is convenient to also use the Riccati parameterization and study the resulting current formula on the normal side of for instance the left NIS interface \cite{Lofwander2003}. This approach is also beneficial when we compare our results with literature where a non-selfconsistent Blonder-Tinkham-Klapwijk (BTK) type of approach has been used \cite{Blonder1982Apr,Tanaka1995Apr}.
Thus lifting the result of Ref.~\cite{Lofwander2003}, the current calculated on the normal side of the left NIS interface at $\xco=0$
can be written as
\begin{align}
j_x(V)\!=\!
2e\NF 
\int\limits_{-\varepsilon_\mathrm{c}}^{\varepsilon_\mathrm{c}} 
&\mathrm{d}\varepsilon
\bigl\langle
v_\mathrm{F}^x
\bigl[
\xx_1( V)
\left\{1 - R_\mathrm{ee} (V) + R_\mathrm{he} (V) \right\}
\nonumber
\\
&+ \xx_2( V)\left\{\bar{T}_\mathrm{he} (V) - \bar{T}_\mathrm{ee} (V) \right\}
\bigr]
\bigr\rangle_{+},
\label{eq:currentFormulaNormalSide}
\end{align}
where the dependencies of all quantities within the brackets on energy $\varepsilon$ and trajectory angle $\varphi_\mathrm{F}$ is understood, while the voltage dependence is emphasized.
Here, $\xx_1$ ($\xx_2)$ is the distribution incoming from the left (right) side of the interface, $\langle \dots\rangle_{+}$ is the average over all momenta where $v_\mathrm{F}^x \equiv v_\mathrm{F} \cos \varphi_\mathrm{F} > 0$.
The remaining four terms are the physical probabilities for normal reflection ($R_\mathrm{ee}$) and Andreev reflection ($R_\mathrm{he}$) for electrons incoming from the left normal metal reservoir, and similarly normal transmission ($\bar{T}_\mathrm{ee}$) and branch-conversion transmission ($\bar{T}_\mathrm{he}$) for electron-like quasiparticles incoming from the superconducting device side. Eq.~\eqref{eq:currentFormulaNormalSide} is a generalization \cite{Lofwander2003} of the well-known BTK current formula for the NIS interface \cite{Blonder1982Apr,Tanaka1995Apr}.
It is a generalization since all four scattering probabilities and both distribution functions, all computed within quasiclassical theory, are voltage dependent. Within traditional BTK theory \cite{Blonder1982Apr} and its generalization to the $d$-wave case \cite{Tanaka1995Apr}, only the incoming distribution from the normal metal side is voltage dependent according to the reservoir and point contact assumptions $\xx_1=\tanh[(\varepsilon-eV)/2\kb\tc]$.
The incoming distribution from the superconducting side, $\xx_2$, as well as the four scattering probabilities in Eq.~\eqref{eq:currentFormulaNormalSide}, are given by their equilibrium values at $V=0$. The conductance within this scattering approach is then simplified to
\begin{align}
G^\mathrm{LB}(V) = 2e\NF
\int\limits_{-\varepsilon_\mathrm{c}}^{\varepsilon_\mathrm{c}} \!&\mathrm{d}\varepsilon
\bigl\langle
v_\mathrm{F}^x
\frac{\mathrm{d}\xx_1(V)}{\mathrm{d}V}
\left[ 1 - R_\mathrm{ee}(0) + R_\mathrm{he}(0)  \right]
\bigr\rangle_+.
\end{align}
In the normal state it reduces to $G_\mathrm{N}=R_\mathrm{N}^{-1}=2e^2\NF v_\mathrm{F}\FSplus{D(\varphi_\mathrm{F})\cos\varphi_\mathrm{F}}$, where $R_\mathrm{N}$ is the normal state interface resistance. We call in the following such an approach the Landauer-B\"uttiker scattering approach, denoted with the superscript LB.

We are now ready to compare our self-consistent stationary non-equilibrium results, blue circles in Fig.~\ref{fig:conductance_comparison}, to the LB scattering approach.
The LB calculations include both the original non-selfconsistent approach \cite{Tanaka1995Apr}, here generalized to include impurity scattering, and a self-consistent calculation of scattering amplitudes for zero voltage where the suppression of the superconducting order parameter is taken into account, as in Ref.~\cite{Lofwander2003}. The self-consistent LB scattering approach (solid green lines in Fig.~\ref{fig:conductance_comparison}) always give a lower conductance compared to the non-selfconsistent LB scattering approach (dashed orange lines in Fig.~\ref{fig:conductance_comparison}) because of the gap suppression near the interface. The conductance in our non-equilibrium calculation (blue filled circles) is similar in shape to the self-consistent LB result, but they do not fully agree.
Corrections to the LB scattering approach are small at small voltages and vanishes at zero voltage where a linear response calculation holds. For intermediate voltages and a high-transparency interface, where the Andreev bound states are substantially broadened, corrections are generally below five percent in the given voltage range, while for low transparency, the difference between the two approaches is more substantial and of order 20 percent due to the voltage dependence of the Andreev state spectral weight, which is also reflected in the scattering amplitudes.
At high voltage, where the $d$-wave order parameter becomes more substantially affected by the non-equilibrium distributions, the corrections grow, until relatively abruptly for these system parameters superconductivity disappears.

\subsection{Subdominant order-parameter component}\label{sec:d_plus_is}

\begin{figure}[t]
    \centering
    \includegraphics{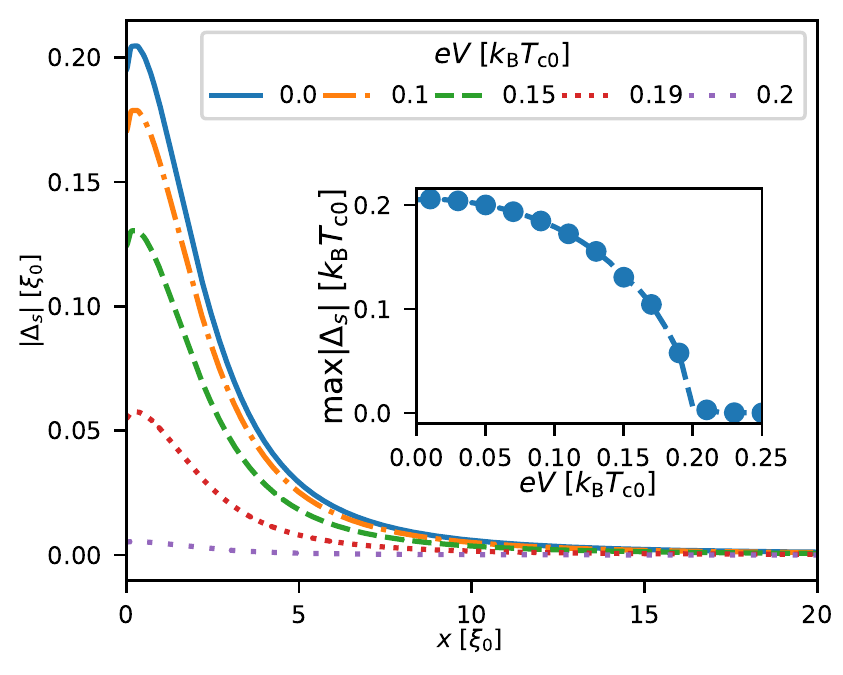}
    \caption{Spatial variation of the absolute value of subdominant component, $|\Delta_\mathrm{s}(\xco)|$, for different voltages. Inset: Largest value of $|\Delta_\mathrm{s}(\xco)|$ as function of bias voltage, where every second datapoint has a marker. Other parameters are $D_0 = 0.5$, $\beta = 1$, $T_\mathrm{s} = 0.2 \tc$, $T = 0.1 \tc $, and $\ell=100\xi_0$ (Born limit).}
    \label{fig:subdominant_voltage_behavior}
\end{figure}

\begin{figure}[t]
    \centering
    \includegraphics{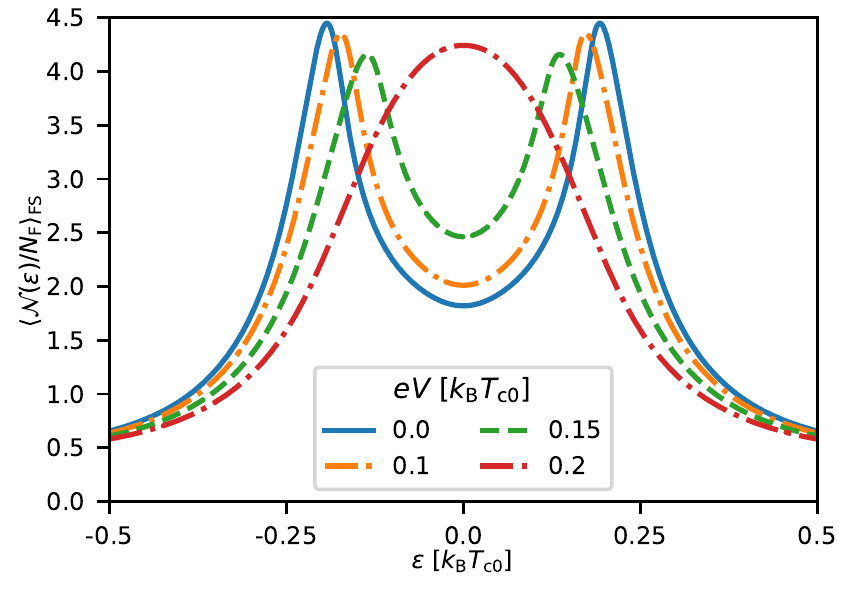}
    \caption{Energy shift of the interface Andreev states by the subdominant order-parameter component as function of applied bias voltage $eV$. Parameters are the same as in Fig.~\ref{fig:subdominant_voltage_behavior} }.
    \label{fig:dos_splitting_swave}
\end{figure}

Even in the absence of an external magnetic field a splitting of the ZBCP has been observed in several experiments at very low temperatures. One mechanism proposed and extensively discussed in the literature \cite{Covington1997Jul,Elhalel2007Mar, Ngai2010Aug} is the presence of a subdominant order parameter of $d_\textrm{xy}$ or $s$ symmetry, in addition to the $d_\mathrm{x^2 -y^2}$ order parameter, in a combination of the form $d_{x^2-y^2}+id_{xy}$ or $d_{x^2-y^2}+is$.
Such combinations result in a superconducting state with broken time-reversal symmetry close to the surface of the superconductor. As a result, the subdominant component becomes enhanced close to the surface and induces a splitting of the ZBCP to finite voltages \cite{Fogelstrom1997Jul}.
Since the $d_{xy}$ subdominant component is very sensitive to impurity scattering, we consider in the following a subdominant $s$-wave order parameter with coupling constant corresponding to a critical temperature of $T_\mathrm{s}=0.2\tc$. This subdominant component is zero in the bulk, $\Delta_\mathrm{s}=0$, but reaches values of $\Delta_\mathrm{s} \sim 0.2 \kb \tc$ close to the surfaces, see the equilibrium shape $\Delta_s (\xco)$ (solid blue line) in Fig.~\ref{fig:subdominant_voltage_behavior}.

Under voltage bias, the injected nonequilibrium distribution
reduces both components of the order parameter.
Since the subdominant component is smaller to begin with, it becomes more strongly suppressed than the dominant component already for small voltages. Fig.~\ref{fig:subdominant_voltage_behavior} shows the suppression of the magnitude of the subdominant component for increasing bias voltage, as computed self-consistently through Eq.~(\ref{eq:DeltaSelfConsistencyEq_hmatrix}). At voltages above $eV \approx 0.2 \kb \tc$ the subdominant component is fully suppressed, as seen in the inset in Fig.~\ref{fig:subdominant_voltage_behavior}. As a result, the Andreev bound states move back towards the Fermi energy for increasing voltage, see Fig.~\ref{fig:dos_splitting_swave}. This shift and the restoration of the zero-energy peak in the surface LDOS with increasing bias voltage also affects the differential conductance.
\begin{figure}[t]
    \centering
    \includegraphics{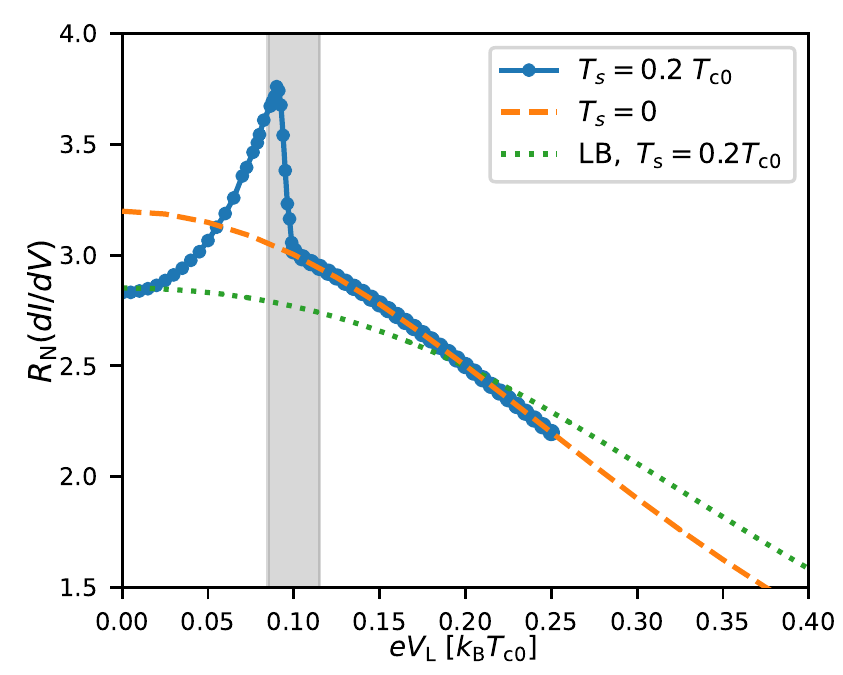}
    \caption{The low-voltage differential conductance, normalized to the normal-state value, for a pure $d$-wave superconductor (dashed orange line) and for the case of an $s$-wave subdominant order parameter with $T_\mathrm{s} = 0.2 \tc$ (solid blue line). Once $eV  = 2 eV_\mathrm{L} \rightarrow 0.2 \kb \tc$ from below, the subdominant component $\Delta_\mathrm{s}$ vanishes, see also the inset of Fig.~\ref{fig:subdominant_voltage_behavior}. The parameters are the same as in Fig.~\ref{fig:subdominant_voltage_behavior}. As comparison we also show results from a Landauer-B\"{u}ttiker scattering-approach (LB) calculation.
   }
    \label{fig:conductance_subdominant}
\end{figure}

As seen in Fig.~\ref{fig:conductance_subdominant}, blue solid line with filled circles, the conductance peak is shifted away from zero voltage due to the equilibrium shift of the Andreev states to finite energy. Once the subdominant component is suppressed at a voltage $eV \approx 0.2 \kb \tc$, the conductance falls back to that of a pure $d$-wave superconductor without any subdominant order parameter component (dashed orange line).
As comparison we have included in Fig.~\ref{fig:conductance_subdominant} the results of a LB calculation (green dotted line).
As seen, the ZBCP is not split and the voltage dependence is smooth and similar to the case without $s$-wave component. This is due to thermal broadening together with the relatively large broadening of the Andreev states by impurity scattering and the coupling to the normal metal through the barrier with transparency $D_0=0.5$. A weak split of the ZBCP appears at very low temperature $T=0.01\tc$ (not shown in the figure), but it is much weaker than our non-equilibrium result at $T=0.1\tc$ (blue curve).
Note that the self-consistent LB approach and our non-equilibrium results coincide at $V=0$ where the linear response approximation holds and all quantities except the voltage perturbation of $\xx_1$ in Eq.~(\ref{eq:currentFormulaNormalSide}) take their $V=0$ equilibrium values.

The main new ingredients in our fully selfconsistent calculation are that also the right-side distribution $\xx_2$, as well as all scattering probabilities, depend on the applied voltage. In contrast to the case without subdominant component considered above (Fig.~\ref{fig:conductance_comparison}), the correction to the LB approach is qualitatively important and result in a dramatic enhancement of the ZBCP split and a rapid drop in the conductance as the $s$-wave component is suppressed by the non-equilibrium distributions. Thermal broadening is not important near the rapid drop as is clearly seen in Fig.~\ref{fig:conductance_subdominant}.
By using the formula in Eq.~(\ref{eq:currentFormulaNormalSide}),
we find that the dominant corrections to the LB scattering approach are those due to self-consistent changes in the amplitudes collected into the factor $[1-R_\mathrm{ee}(V) + R_\mathrm{he}(V)]$ in Eq.~\eqref{eq:currentFormulaNormalSide}. The applied bias voltage leads to a quick suppression of the subdominant order parameter component $\Delta_\mathrm{s}$ through the highly non-equilibrium forms of the distribution functions. As a consequence, the surface Andreev bound states move back to zero energy, as seen in Fig.~\ref{fig:dos_splitting_swave}. The resonances in the scattering amplitudes mirror this spectral rearrangement. As a result the term proportional to the voltage derivative $\mathrm{d}[1 - R_\mathrm{ee}(V) + R_\mathrm{he}(V)]/\mathrm{d}V$ also gives a large contribution of roughly 20 percent of the total conductance. In contrast, the backflow contribution proportional to $\xx_2$ in Eq.~\eqref{eq:currentFormulaNormalSide} leads to small corrections of less than five percent. At a critical voltage $\sim 0.2\kb\tc$, the subdominant component is fully suppressed and a rapid drop of the conductance back to the pure $d$-wave curve, the orange dashed line in Fig.~\ref{fig:conductance_subdominant}), is visible.
We conclude that the voltage dependence of scattering probabilities can be of great importance when the non-equilibrium distributions couple back through the order parameter self-consistency.

\section{Summary}\label{sec:summary}

In summary we have presented a comprehensive study of the stationary non-equilibrium response of a $d$-wave superconductor coupled to two normal-metal reservoirs under a voltage bias, taking into account different orientations of the order parameter relative to the interfaces to the reservoirs, formation of interface zero-energy Andreev bound states, scalar impurity scattering, and a possible $s$-wave subdominant component of the order parameter. In all cases we ensure current conservation by computing all self-energies selfconsistently taking into account the non-equilibrium distribution functions. For the case of a pure $d$-wave order parameter, we have found that charge imbalance extends into the bulk of the superconductor due to the presence of the nodes of the $d$-wave order parameter and the pair breaking effects of scalar impurities. Charge imbalance is enhanced for orientations where zero-energy interface Andreev bound states are formed and it is also enhanced for unitary limit impurity scattering which induce a low-energy band of quasiparticles states. The non-equilibrium distribution function induced in the superconductor suppresses the $d$-wave order parameter and superconductivity disappears for voltages approaching the gap voltage. Nevertheless, despite the induced non-equilibrium state, the resulting conductance-voltage dependencies for the case of pure $d_{x^2-y^2}$ superconducting order resembles the results of a non-selfconsistent Landauer-B\"uttiker approach, if the original approach \cite{Tanaka1995Apr} is corrected for the zero-voltage ($V=0$) equilibrium suppression of the order parameter near the interfaces and the broadening effect of impurities. This holds until superconductivity is suppressed at voltages approaching the gap voltage.

When we allow for a subdominant component with $s$-wave symmetry, forming the time-reversal symmetry breaking combination $d_{x^2-y^2}+is$ near the interfaces, the effects of the non-equilibrium distribution is much more severe. In this case, the zero-bias conductance peak split is dramatically enhanced compared with the non-selfconsistent Landauer-B\"uttiker approach. This effect is due to the suppression of the subdominant $s$-wave order parameter when the voltage is applied. This leads to spectral rearrangements and corrections to the scattering amplitudes. As a result, a non-thermally broadened split of the zero-bias conductance peak appears.

The Landauer-B\"uttiker approach holds for point-contact geometries, where the contact radius is smaller than the coherence length $\xi_0$. Since $\xi_0$ is small in high-temperature superconductors such contacts are challenging to fabricate. Our approach is applicable to the complementary geometry with very wide and transparent contacts where the induced non-equilibrium distribution is important. Such contacts to superconducting films are experimentally feasible \cite{Baghdadi2015Jul,Baghdadi2017} and could serve as an interesting probe of the rich surface physics of unconventional superconductors, including the possibility of subdominant order parameters and time-reversal symmetry breaking.

\begin{acknowledgments}
We thank F. Lombardi for valuable discussion.
We acknowledge the Swedish research council for financial support. The computations were enabled by resources provided by the Swedish National Infrastructure for Computing (SNIC) at NSC partially funded by the Swedish Research Council through grant agreement no. 2018-05973.
\end{acknowledgments}

\appendix

\section{Quasiclassical theory}

\label{sec:Appendix}
Here, we review the key set of equations of the quasiclassical theory of superconductivity that we base our calculations on. For further details we refer to Refs.~\cite{Seja2021Sep,Seja2022Mar}.

In the stationary non-equilibrium case under consideration we need to calculate the retarded and advanced Green's functions as well as the Keldysh component. They are organized into a matrix in Keldysh space as
\begin{equation}
\check{g}(\pF,\RR,\varepsilon) =
\begin{pmatrix}
\hat{g}^\mathrm{R}(\pF,\RR,\varepsilon) & \hat{g}^\mathrm{K}(\pF,\RR,\varepsilon)
\\
0 & \hat{g}^\mathrm{A}(\pF,\RR,\varepsilon)
\end{pmatrix}.
\label{eq:gcheckDefinition}
\end{equation}
For brevity we often drop the explicit references to the dependencies on momentum direction $\pF$, coordinate $\RR$, and energy $\varepsilon$. 
All three elements in Eq.~\eqref{eq:gcheckDefinition} are matrices in Nambu space, as indicated by the $\hat{~}$.
In the steady-state, $\check{g}$ is time-independent and obeys the so-called Eilenberger equation,
\begin{equation}
i \hbar \vF \cdot \nabla \check{g} + \left[ \varepsilon \hat{\tau}_3 \check{1} - \check{h}, \check{g} \right] = 0,
\label{eq:transportequation}
\end{equation}
where $[\check A,\check B]$ denotes a commutator between matrices $\check A$ and $\check B$,
as well as the normalization condition $\check{g}^2 = - \pi^2 \check{1}$. These two equations were first derived by 
Eilenberger \cite{Eilenberger1968Apr}, and separately Larkin and Ovchinnikov \cite{Larkin1969}, and the generalization to nonequilibrium is due to Eliashberg \cite{Eliashberg1971}.
The quasiclassical Hamiltonian $\check{h}$ has the same Keldysh structure as the propagator in Eq.~(\ref{eq:gcheckDefinition}). The components in Nambu space are
\begin{align}
    \hat{h}_\mathrm{}^\mathrm{R,A}(\RR,\varepsilon) =
\begin{pmatrix}
\Sigma^\mathrm{R,A}(\RR,\varepsilon) & \Delta^\mathrm{R,A}(\RR,\varepsilon)
\\
\tilde\Delta^\mathrm{R,A}(\RR,\varepsilon) & \tilde\Sigma^\mathrm{R,A}(\RR,\varepsilon)
\end{pmatrix},
\label{eq:ImpRetarded}
\end{align}
and
\begin{align}
\hat{h}^\mathrm{K}(\RR,\varepsilon) \equiv
\begin{pmatrix}
\Sigma^\mathrm{K}(\RR,\varepsilon) & \Delta^\mathrm{K}(\RR,\varepsilon)
\\
- \tilde{\Delta}^\mathrm{K}(\RR,\varepsilon) & - \tilde{\Sigma}^\mathrm{K}(\RR,\varepsilon)
\end{pmatrix}
.
\end{align}
These self-energies are defined below.

The elements of $\check{g}$ can be parametrized using coherence amplitudes $\gamma, \tilde{\gamma}$ \cite{Nagato1993, Schopohl1995, Schopohl1998} and generalized distribution functions $x, \tilde{x}$ \cite{eschrig_distribution_2000,Eschrig2009Oct}. Using the definitions of $\mathcal{G}^R \equiv ( 1 - \gamma^\mathrm{R} \tilde{\gamma}^\mathrm{R})^{-1}$ and $\mathcal{F}^\mathrm{R} \equiv \mathcal{G}^\mathrm{R}\gamma^\mathrm{R}$, we have for the retarded element
\begin{align}
\hat{g}^\mathrm{R} = 
-2\pi i
\!
\begin{pmatrix}
\mathcal{G}& \mathcal{F}
 \\
-\tilde{\mathcal{F}} & -\tilde{\mathcal{G}}
\end{pmatrix}^\mathrm{R}
+ i \pi \hat{\tau}_3,
\end{align}
and an analogous expression for the advanced function $\hat{g}^A$. The Keldysh component reads
\begin{align}
\hat{g}^\mathrm{K}
&= 
\begin{pmatrix}
g& f
\\
- \tilde{f} & - \tilde{g}
\end{pmatrix}^\mathrm{K} 
=
 - 2 \pi i 
\begin{pmatrix}
\mathcal{X} & \mathcal{Y}
\\
\tilde{\mathcal{Y}} & \tilde{\mathcal{X}}
\end{pmatrix}^\mathrm{K}.
\nonumber
\\
&
\equiv
-2 \pi i 
\begin{pmatrix}
\mathcal{G} & \mathcal{F}
\\
-\tilde{\mathcal{F}} & -\tilde{\mathcal{G}}
\end{pmatrix}^\mathrm{R}
\begin{pmatrix}
\xx & 0
\\ 0 & \tilde{\xx}
\end{pmatrix}
\begin{pmatrix}
\mathcal{G} & \mathcal{F}
\\
-\tilde{\mathcal{F}} & -\tilde{\mathcal{G}}
\end{pmatrix}^\mathrm{A}.
\label{eq:riccati_x}
\end{align}
An important symmetry in the theory is particle-hole conjugation, which can be expressed as
\begin{align}
\tilde{A}(\pF,\RR,\varepsilon) = A^*(-\pF,\RR,-\varepsilon^*).
\label{eq:TildeSymmetry}
\end{align}
A set of coupled transport equations for the parametrizing functions can be derived from Eq.~\eqref{eq:transportequation}. The Riccati equation for the coherence amplitude reads
\begin{align}
\left( i\hbar\vF\cdot\nabla + 2 \varepsilon \right) \gamma^{\mathrm{R},\mathrm{A}}
\!=\!\bigl( \gamma \tilde{\Delta} \gamma + \Sigma \gamma - \gamma \tilde{\Sigma} - \Delta \bigr)^{\mathrm{R},\mathrm{A}}\!,
\label{eq:GammaEquation}
\end{align}
while the equation for the distribution function $\xx$ is
\begin{align}
&i \hbar \vF \cdot \nabla \xx - \left[ \gamma \tilde{\Delta} + \Sigma \right]^\mathrm{R} \xx - \xx \left[ \Delta \tilde{\gamma} - \Sigma \right]^\mathrm{A}
\nonumber
\\
&=
- \gamma^\mathrm{R} \tilde{\Sigma}^\mathrm{K} \tilde{\gamma}^\mathrm{A} + \Delta^\mathrm{K} \tilde{\gamma}^\mathrm{A} + \gamma^\mathrm{R} \tilde{\Delta}^\mathrm{K} - \Sigma^\mathrm{K}.
\label{eq:xEquation}
\end{align}
Application of Eq.~\eqref{eq:TildeSymmetry} to these two equations gives the corresponding equations for $\tilde{\gamma}^\mathrm{R,A}$ and $\tilde{\xx}$.
Solutions to both transport equations are found by propagating from a start point to an end point on a trajectory determined by the Fermi momentum $\vF$, in our case fully determined by the momentum orientation angle $\varphi_\mathrm{F}$. 
At interfaces between the superconductor and the normal-metal reservoirs, both the coherence and distribution functions have to be connected by boundary conditions. The latter are derived from the scattering matrices for an insulating barrier between a normal metal and a superconductor, see Refs.~\onlinecite{eschrig_distribution_2000,Eschrig2009Oct, Zhao2004Oct}. 

Having obtained solutions to Eqs.~\eqref{eq:GammaEquation}-\eqref{eq:xEquation}, we can update all self-energies. They consist of the mean field order parameter and scalar impurity self energy as
\begin{equation}
\check{h}(\pF,\RR,\varepsilon) = \check{h}_\mathrm{mf}(\pF,\RR) + \check{h}_\mathrm{s}(\RR,\varepsilon).
\end{equation}
The Keldysh matrix structure of the mean field order parameter is simple,
$\check{h}_\mathrm{mf} = \hat\Delta\check{1}$,
while the Nambu structure is $\hat\Delta=\Re(\Delta)\hat\tau_1-\Im(\Delta)\hat\tau_2$. The order parameter consists of a sum over the $d$-wave and subdominant $s$-wave components. Each component satisfies a gap equation
\begin{equation}
\Delta_\Gamma(\RR)\!=\!\NF\lambda_\Gamma\!\int\limits_{-\varepsilon_\mathrm{c}}^{\varepsilon_c}\!\frac{d\varepsilon}{8\pi i}
\FS{\mbox{Tr}\left[i\sigma_2\eta_\Gamma(\pF) f^\mathrm{K}(\pF,\RR,\varepsilon)\right]},
\label{eq:gapequation}
\end{equation}
where $\Gamma=\mathrm{d},\,\mathrm{s}$. The basis functions are $\eta_\mathrm{d}(\pF)=\sqrt{2}\cos[2(\varphi_\mathrm{F}-\alpha)]$ and $\eta_\mathrm{s}(\pF)=1$. The pairing interactions $\lambda_\Gamma$ are eliminated in favor of the clean-limit critical temperatures for each channel by the replacement
\begin{equation}
\frac{1}{\NF\lambda_\Gamma}\rightarrow \ln\left(\frac{T}{T_{\mathrm{c}\Gamma}}\right)
+\int^{\varepsilon_\mathrm{c}}_{\varepsilon_\mathrm{c}}
\frac{d\varepsilon}{2}
\frac{1}{{\varepsilon}}
\tanh\left(\frac{\varepsilon}{2\kb T}\right).
\end{equation}
In the main text, since the $d$-wave is the dominant component, we let $\tc=T_\mathrm{cd}$ and use $\kb\tc$ as the natural energy scale. The critical temperature of the subdominant $s$-wave is for short denoted $T_\mathrm{s}<\tc$.

The impurity self-energies are calculated in the non-crossing approximation for the $t$-matrix equation \cite{AGD:Book}, discussed in detail in Ref.~\onlinecite{Seja2022Mar}. The matrix structure is
\begin{align}
\check{h}_s = n_\mathrm{i} \check{t} \equiv n_\mathrm{i}
\begin{pmatrix}
\hat{t}^\mathrm{\,R} & \hat{t}^\mathrm{\,K}
\\
0 & \hat{t}^\mathrm{\,A}
\end{pmatrix}.
\label{eq:selfEnergyTMatrixEquation}
\end{align}
For scattering that is isotropic in momentum space with an $s$-wave scattering potential $u_0$ the elements of $\check{t}$ satisfy the equations
\begin{align}
\hat{t}^\mathrm{\,R,A} &= \frac{u_0 \hat{1} + u_0^2 \NF \FS{\hat{g}^\mathrm{R,A}}}{\hat{1} - \left[  u_0 \NF \FS{\hat{g}^\mathrm{R,A} }\right]^2   },
\label{eq:tRDefinition}
\\
\hat{t}^\mathrm{\,K} &= \NF \hat{t}^\mathrm{\,R} \FS{\hat{g}^\mathrm{K}} \hat{t}^\mathrm{\,A}.
\label{eq:tKDefinition}
\end{align}

This procedure of solving Eqs.~\eqref{eq:GammaEquation}-\eqref{eq:xEquation} and updating the selfenergies is iterated until a fully self-consistent solution is found and current is conserved throughout the structure. In the present paper, we allow for a maximum relative error $|j(x) - j_\mathrm{I}|/|j_\mathrm{I}| < 5\cdot10^{-3}$, so the current everywhere deviates less than half a percent from $j_\mathrm{I}$, the current at the interface to the normal-metal reservoirs. 
The charge current is determined by the Keldysh component $\hat{g}^K$, 
\begin{align}
\mathbf{j}(\RR) &= e\NF\!\int\limits_{-\infty}^{\infty}\!
\frac{\mathrm{d}\varepsilon}{8\pi i} 
\FS{\mathrm{Tr}\left[ \vF \hat{\tau}_3 \hat{g}^\mathrm{K}(\pF,\RR,\varepsilon) \right]},
\label{eq:ChargeCurrentDefinition}
\end{align}
as is the local electrochemical potential
\begin{equation}
\phi(\RR) = \frac{1}{2e} \int\limits_{-\infty}^{\infty} \frac{\mathrm{d}\varepsilon}{8 \pi i} \FS{\mathrm{Tr}~\hat{g}^\mathrm{K}(\pF,\RR,\varepsilon)}.
\label{eq:phi}
\end{equation}
The retarded component $\hat{g}^R$ determines the normalized, momentum-resolved local density of states per spin
\begin{equation}
\mathcal{N}(\pF, \RR, \varepsilon) = -\frac{1}{4\pi}\mathrm{Im}~\mathrm{Tr}\left[\hat{\tau}_3 \hat{g}^R(\pF, \RR, \varepsilon)\right],
\end{equation}
and thus the full density of states
\begin{equation}
N(\RR,\varepsilon) = 2\NF\FS{\mathcal{N}(\pF, \RR, \varepsilon)}.
\label{eq:DensityOfStates_averaged}
\end{equation}
The Keldysh component $\hat{g}^\mathrm{K}$ can, in addition to Eq.~\eqref{eq:riccati_x}, also be parametrized as
\begin{align}
\hat{g}^\mathrm{K} = \hat{g}^\mathrm{R} \hat{f}  - \hat{f} \hat{g}^\mathrm{A},
\label{eq:gKeldyshInH}
\end{align}
where the distribution matrix $\hat{f}$ reads
\begin{align}
\hat{f} =
\begin{pmatrix}
h & 0 
\\
0 & -\tilde{h}
\end{pmatrix}
= f_1 \hat{1} + f_3 \hat{\tau}_3.
\label{eq:h-splitting}
\end{align}
The non-equilibrium electron distribution function is then given by
\begin{align}
\fe = \tfrac{1}{2}(1 - h) = \tfrac{1}{2}\left(1 - f_1 - f_3\right),
\end{align}
which becomes a Fermi-Dirac distribution function in equilibrium.
We refer to $f_1$ ($f_3$) as the energy-like (charge-like) mode, connecting to the nomenclature established for diffusive superconductors \cite{Schmid1975Jul}. After Fermi-surface averaging, they have different symmetries as function of energy, namely
\begin{align}
\FS{f_1}(\varepsilon) = - \FS{f_1}(-\varepsilon),
\label{eq:energyModeSymmetry}
\\
\FS{f_3}(\varepsilon) = \FS{f_3}(-\varepsilon)
\label{eq:chargeModeSymmetry}
\end{align}
which corresponds to the symmetries obtained in the diffusive case.
The function $h$ can be related to $\xx$ by
\begin{align}
\xx = h + \gamma^\mathrm{R} \tilde{h} \tilde{\gamma}^\mathrm{A}.
\label{eq:xAndhRelation}
\end{align}
The equation of motion for $\xx$ and $\tilde{\xx}$ is easier to solve, while the final results can be easier to interpret in terms of $h$, or $f_{\mathrm{e}/1/3}$. As an example, we can write the gap equations in Eq.~\eqref{eq:gapequation} as 
\begin{equation}
\frac{\Delta_\Gamma(\RR)}{\NF\lambda_\Gamma/2}\! =\! \!\int\limits_{-\varepsilon_\mathrm{c}}^{\varepsilon_\mathrm{c}}\!\mathrm{d}\varepsilon
\FS{\eta_\Gamma(\pF)
\left[ f_1( \mathcal{F}^R + \mathcal{F}^A) - f_3( \mathcal{F}^R - \mathcal{F}^A) \right]},
\label{eq:DeltaSelfConsistencyEq_hmatrix}
\end{equation}
and the charge current, Eq.~\eqref{eq:ChargeCurrentDefinition}, as 
\begin{align}
\mathbf{j}(\RR)\!=\!-e\NF \int\limits_{-\infty}^{\infty}
\mathrm{d}\varepsilon
\FS{ \vF f_1(\pF,\RR,\varepsilon) \mathcal{N}(\pF,\RR,\varepsilon) }.
\label{eq:ChargeCurrentDefinition_hmatrix}
\end{align}
The natural unit for charge current is then $j_0=e v_\mathrm{F}\NF\kb\tc$. The electrochemical potential, Eq.~\eqref{eq:phi}, can similarly be written as 
\begin{equation}
\phi(\RR) = -\frac{1}{2e} \int\limits_{-\infty}^{\infty} \mathrm{d}\varepsilon \FS{f_3(\pF,\RR,\varepsilon) \mathcal{N}(\pF,\RR,\varepsilon)}.
\label{eq:phi_hmatrix}
\end{equation}

The distribution function $h$ can be split into a local-equilibrium function $h^\mathrm{le}$, and an anomalous function $h^\mathrm{a}$,
\begin{align}
h = h^\mathrm{le} + ( h - h^\mathrm{le}) \equiv h^\mathrm{le} + h^a,
\label{eq:hDistributionSplitting}
\end{align}
where $h^\mathrm{le}$ is given by
\begin{equation}
h^\mathrm{le}(\RR,\varepsilon) = \tanh\frac{\varepsilon-e\phi(\RR)}{2T}.
\label{eq:h_le}
\end{equation}
The splitting introduced in Eq.~\eqref{eq:hDistributionSplitting} directly translate to an analogous splitting of the function $\xx$ by using Eq.~\eqref{eq:xAndhRelation}, and for the two modes $f_1$ and $f_3$ by using Eq.~\eqref{eq:h-splitting}. As a consequence observables such as the charge current can be similarly split,
\begin{equation}
\mathbf{j}(\RR) = \mathbf{j}^\mathrm{le}(\RR) + \mathbf{j}^\mathrm{a}(\RR).
\label{eq:CurrentSplitting}
\end{equation}
Since $h^\mathrm{le}$, and hence $f_\mathrm{1,3}^\mathrm{le}$, are independent of $\pF$, Eq.~\eqref{eq:ChargeCurrentDefinition_hmatrix} shows that the local-equilibrium current $\mathbf{j}^\mathrm{le}$ can only be non-zero if there is an asymmetry in the angle-resolved spectrum $\mathcal{N}(\pF,\RR,\varepsilon)$. In our case, this asymmetry is created by the self-consistently determined superflow $\mathbf{p}_\mathrm{s}$ in the superconductor. The anomalous current $\mathbf{j}^\mathrm{a}$, due to the distribution $h^\mathrm{a}$, then contains all current flow due to differences in occupation of left-moving and right-moving quasiparticles. 
\pagebreak
A measure of this difference in occupation are the right-mover (left-mover) quasipotential $\phi_+$ ($\phi_-$), defined as
\begin{equation}
\phi_{\pm} := \phi -\frac{1}{2e}\int\limits_{-\infty}^{\infty}\! \frac{d\varepsilon}{2}~ \left( \langle\mathcal{X}^\mathrm{a} \rangle_{\pm}  + \langle \tilde{\mathcal{X} }^\mathrm{a} \rangle_\mp \right),
\label{eq:PhiLeftRight}
\end{equation}
where $\mathcal{X}^a$ is obtained by using $\xx^\mathrm{a}$ in Eq.~\eqref{eq:riccati_x}. In Eq.~\eqref{eq:PhiLeftRight}, the $\langle \dots \rangle_\pm$ denotes a partial Fermi-surface average
\begin{align}
\langle A \rangle_\pm &\equiv 
\int\limits_{0}^{2\pi}\!
\frac{d\varphi_\mathrm{F}}{\pi}
\!~ A(\varphi_\mathrm{F}) \Theta(\pm \cos \varphi_\mathrm{F}),
\label{eq:PartialFermiAverage}
\end{align}
where the Heaviside step function $\Theta(\pm\cos\varphi_\mathrm{F})$ gives unity if $v_\mathrm{F}^x$, as defined in Eq.~\eqref{eq:vF_x_definition}, is positive (+) or negative (-), and zero otherwise. 
In the normal state, the left-mover and right-mover quasipotentials satisfy 
$\phi = (\phi_+ + \phi_-)/2$,
which connects to the concepts used in normal-state ballistic systems \cite{Datta2017Feb}.


\begin{thebibliography}{65}%
\makeatletter
\providecommand \@ifxundefined [1]{%
 \@ifx{#1\undefined}
}%
\providecommand \@ifnum [1]{%
 \ifnum #1\expandafter \@firstoftwo
 \else \expandafter \@secondoftwo
 \fi
}%
\providecommand \@ifx [1]{%
 \ifx #1\expandafter \@firstoftwo
 \else \expandafter \@secondoftwo
 \fi
}%
\providecommand \natexlab [1]{#1}%
\providecommand \enquote  [1]{``#1''}%
\providecommand \bibnamefont  [1]{#1}%
\providecommand \bibfnamefont [1]{#1}%
\providecommand \citenamefont [1]{#1}%
\providecommand \href@noop [0]{\@secondoftwo}%
\providecommand \href [0]{\begingroup \@sanitize@url \@href}%
\providecommand \@href[1]{\@@startlink{#1}\@@href}%
\providecommand \@@href[1]{\endgroup#1\@@endlink}%
\providecommand \@sanitize@url [0]{\catcode `\\12\catcode `\$12\catcode
  `\&12\catcode `\#12\catcode `\^12\catcode `\_12\catcode `\%12\relax}%
\providecommand \@@startlink[1]{}%
\providecommand \@@endlink[0]{}%
\providecommand \url  [0]{\begingroup\@sanitize@url \@url }%
\providecommand \@url [1]{\endgroup\@href {#1}{\urlprefix }}%
\providecommand \urlprefix  [0]{URL }%
\providecommand \Eprint [0]{\href }%
\providecommand \doibase [0]{https://doi.org/}%
\providecommand \selectlanguage [0]{\@gobble}%
\providecommand \bibinfo  [0]{\@secondoftwo}%
\providecommand \bibfield  [0]{\@secondoftwo}%
\providecommand \translation [1]{[#1]}%
\providecommand \BibitemOpen [0]{}%
\providecommand \bibitemStop [0]{}%
\providecommand \bibitemNoStop [0]{.\EOS\space}%
\providecommand \EOS [0]{\spacefactor3000\relax}%
\providecommand \BibitemShut  [1]{\csname bibitem#1\endcsname}%
\let\auto@bib@innerbib\@empty
\bibitem [{\citenamefont {Giaever}(1960)}]{Giaever1960}%
  \BibitemOpen
  \bibfield  {author} {\bibinfo {author} {\bibfnamefont {I.}~\bibnamefont
  {Giaever}},\ }\bibfield  {title} {\bibinfo {title} {Energy gap in
  superconductors measured by electron tunneling},\ }\href
  {https://doi.org/10.1103/PhysRevLett.5.147} {\bibfield  {journal} {\bibinfo
  {journal} {Phys. Rev. Lett.}\ }\textbf {\bibinfo {volume} {5}},\ \bibinfo
  {pages} {147} (\bibinfo {year} {1960})}\BibitemShut {NoStop}%
\bibitem [{\citenamefont {Bardeen}(1961)}]{Bardeen1961}%
  \BibitemOpen
  \bibfield  {author} {\bibinfo {author} {\bibfnamefont {J.}~\bibnamefont
  {Bardeen}},\ }\bibfield  {title} {\bibinfo {title} {Tunnelling from a
  many-particle point of view},\ }\href
  {https://doi.org/10.1103/PhysRevLett.6.57} {\bibfield  {journal} {\bibinfo
  {journal} {Phys. Rev. Lett.}\ }\textbf {\bibinfo {volume} {6}},\ \bibinfo
  {pages} {57} (\bibinfo {year} {1961})}\BibitemShut {NoStop}%
\bibitem [{\citenamefont {Schrieffer}(1999)}]{Schrieffer:Book}%
  \BibitemOpen
  \bibfield  {author} {\bibinfo {author} {\bibfnamefont {J.~R.}\ \bibnamefont
  {Schrieffer}},\ }\href@noop {} {\emph {\bibinfo {title} {Theory of
  Superconductivity}}},\ Advanced Book Classics\ (\bibinfo  {publisher}
  {Perseus Books},\ \bibinfo {address} {Reading, Massachusetts, USA},\ \bibinfo
  {year} {1999})\BibitemShut {NoStop}%
\bibitem [{\citenamefont {Blonder}\ \emph {et~al.}(1982)\citenamefont
  {Blonder}, \citenamefont {Tinkham},\ and\ \citenamefont
  {Klapwijk}}]{Blonder1982Apr}%
  \BibitemOpen
  \bibfield  {author} {\bibinfo {author} {\bibfnamefont {G.~E.}\ \bibnamefont
  {Blonder}}, \bibinfo {author} {\bibfnamefont {M.}~\bibnamefont {Tinkham}},\
  and\ \bibinfo {author} {\bibfnamefont {T.~M.}\ \bibnamefont {Klapwijk}},\
  }\bibfield  {title} {\bibinfo {title} {{Transition from metallic to tunneling
  regimes in superconducting microconstrictions: Excess current, charge
  imbalance, and supercurrent conversion}},\ }\href
  {https://doi.org/10.1103/PhysRevB.25.4515} {\bibfield  {journal} {\bibinfo
  {journal} {Phys. Rev. B}\ }\textbf {\bibinfo {volume} {25}},\ \bibinfo
  {pages} {4515} (\bibinfo {year} {1982})}\BibitemShut {NoStop}%
\bibitem [{\citenamefont {Meservey}\ and\ \citenamefont
  {Tedrow}(1994)}]{Merservey1994}%
  \BibitemOpen
  \bibfield  {author} {\bibinfo {author} {\bibfnamefont {R.}~\bibnamefont
  {Meservey}}\ and\ \bibinfo {author} {\bibfnamefont {P.}~\bibnamefont
  {Tedrow}},\ }\bibfield  {title} {\bibinfo {title} {Spin-polarized electron
  tunneling},\ }\href
  {https://doi.org/https://doi.org/10.1016/0370-1573(94)90105-8} {\bibfield
  {journal} {\bibinfo  {journal} {Physics Reports}\ }\textbf {\bibinfo {volume}
  {238}},\ \bibinfo {pages} {173} (\bibinfo {year} {1994})}\BibitemShut
  {NoStop}%
\bibitem [{\citenamefont {Soulen}\ \emph {et~al.}(1998)\citenamefont {Soulen},
  \citenamefont {Byers}, \citenamefont {Osofsky}, \citenamefont {Nadgorny},
  \citenamefont {Ambrose}, \citenamefont {Cheng}, \citenamefont {Broussard},
  \citenamefont {Tanaka}, \citenamefont {Nowak}, \citenamefont {Moodera},
  \citenamefont {Barry},\ and\ \citenamefont {Coey}}]{Soulen1998}%
  \BibitemOpen
  \bibfield  {author} {\bibinfo {author} {\bibfnamefont {R.~J.}\ \bibnamefont
  {Soulen}}, \bibinfo {author} {\bibfnamefont {J.~M.}\ \bibnamefont {Byers}},
  \bibinfo {author} {\bibfnamefont {M.~S.}\ \bibnamefont {Osofsky}}, \bibinfo
  {author} {\bibfnamefont {B.}~\bibnamefont {Nadgorny}}, \bibinfo {author}
  {\bibfnamefont {T.}~\bibnamefont {Ambrose}}, \bibinfo {author} {\bibfnamefont
  {S.~F.}\ \bibnamefont {Cheng}}, \bibinfo {author} {\bibfnamefont {P.~R.}\
  \bibnamefont {Broussard}}, \bibinfo {author} {\bibfnamefont {C.~T.}\
  \bibnamefont {Tanaka}}, \bibinfo {author} {\bibfnamefont {J.}~\bibnamefont
  {Nowak}}, \bibinfo {author} {\bibfnamefont {J.~S.}\ \bibnamefont {Moodera}},
  \bibinfo {author} {\bibfnamefont {A.}~\bibnamefont {Barry}},\ and\ \bibinfo
  {author} {\bibfnamefont {J.~M.~D.}\ \bibnamefont {Coey}},\ }\bibfield
  {title} {\bibinfo {title} {Measuring the spin polarization of a metal with a
  superconducting point contact},\ }\href
  {https://doi.org/10.1126/science.282.5386.85} {\bibfield  {journal} {\bibinfo
   {journal} {Science}\ }\textbf {\bibinfo {volume} {282}},\ \bibinfo {pages}
  {85} (\bibinfo {year} {1998})}\BibitemShut {NoStop}%
\bibitem [{\citenamefont {Upadhyay}\ \emph {et~al.}(1998)\citenamefont
  {Upadhyay}, \citenamefont {Palanisami}, \citenamefont {Louie},\ and\
  \citenamefont {Buhrman}}]{Upadhyay1998}%
  \BibitemOpen
  \bibfield  {author} {\bibinfo {author} {\bibfnamefont {S.~K.}\ \bibnamefont
  {Upadhyay}}, \bibinfo {author} {\bibfnamefont {A.}~\bibnamefont
  {Palanisami}}, \bibinfo {author} {\bibfnamefont {R.~N.}\ \bibnamefont
  {Louie}},\ and\ \bibinfo {author} {\bibfnamefont {R.~A.}\ \bibnamefont
  {Buhrman}},\ }\bibfield  {title} {\bibinfo {title} {{Probing Ferromagnets
  with Andreev Reflection}},\ }\href
  {https://doi.org/10.1103/PhysRevLett.81.3247} {\bibfield  {journal} {\bibinfo
   {journal} {Phys. Rev. Lett.}\ }\textbf {\bibinfo {volume} {81}},\ \bibinfo
  {pages} {3247} (\bibinfo {year} {1998})}\BibitemShut {NoStop}%
\bibitem [{\citenamefont {Geerk}\ \emph {et~al.}(1988)\citenamefont {Geerk},
  \citenamefont {Xi},\ and\ \citenamefont {Linker}}]{Geerk1988}%
  \BibitemOpen
  \bibfield  {author} {\bibinfo {author} {\bibfnamefont {J.}~\bibnamefont
  {Geerk}}, \bibinfo {author} {\bibfnamefont {X.~X.}\ \bibnamefont {Xi}},\ and\
  \bibinfo {author} {\bibfnamefont {G.}~\bibnamefont {Linker}},\ }\bibfield
  {title} {\bibinfo {title} {Electron tunneling into thin films of y1ba2cu3o7
  josephson effect through
  ${\mathrm{yba}}_{2}{\mathrm{cu}}_{3}{\mathrm{o}}_{7}$},\ }\href
  {https://doi.org/10.1007/BF01314271} {\bibfield  {journal} {\bibinfo
  {journal} {Z. Phys. B: Cond. Mat.}\ }\textbf {\bibinfo {volume} {73}},\
  \bibinfo {pages} {329} (\bibinfo {year} {1988})}\BibitemShut {NoStop}%
\bibitem [{\citenamefont {Hu}(1994)}]{Hu1994}%
  \BibitemOpen
  \bibfield  {author} {\bibinfo {author} {\bibfnamefont {C.-R.}\ \bibnamefont
  {Hu}},\ }\bibfield  {title} {\bibinfo {title} {Midgap surface states as a
  novel signature for
  ${\mathit{d}}_{\mathit{x}\mathit{a}}^{2}$-${\mathit{x}}_{\mathit{b}}^{2}$-wave
  superconductivity},\ }\href {https://doi.org/10.1103/PhysRevLett.72.1526}
  {\bibfield  {journal} {\bibinfo  {journal} {Phys. Rev. Lett.}\ }\textbf
  {\bibinfo {volume} {72}},\ \bibinfo {pages} {1526} (\bibinfo {year}
  {1994})}\BibitemShut {NoStop}%
\bibitem [{\citenamefont {Tanaka}\ and\ \citenamefont
  {Kashiwaya}(1995)}]{Tanaka1995Apr}%
  \BibitemOpen
  \bibfield  {author} {\bibinfo {author} {\bibfnamefont {Y.}~\bibnamefont
  {Tanaka}}\ and\ \bibinfo {author} {\bibfnamefont {S.}~\bibnamefont
  {Kashiwaya}},\ }\bibfield  {title} {\bibinfo {title} {{Theory of Tunneling
  Spectroscopy of $\mathit{d}$-Wave Superconductors}},\ }\href
  {https://doi.org/10.1103/PhysRevLett.74.3451} {\bibfield  {journal} {\bibinfo
   {journal} {Phys. Rev. Lett.}\ }\textbf {\bibinfo {volume} {74}},\ \bibinfo
  {pages} {3451} (\bibinfo {year} {1995})}\BibitemShut {NoStop}%
\bibitem [{\citenamefont {Covington}\ \emph {et~al.}(1997)\citenamefont
  {Covington}, \citenamefont {Aprili}, \citenamefont {Paraoanu}, \citenamefont
  {Greene}, \citenamefont {Xu}, \citenamefont {Zhu},\ and\ \citenamefont
  {Mirkin}}]{Covington1997Jul}%
  \BibitemOpen
  \bibfield  {author} {\bibinfo {author} {\bibfnamefont {M.}~\bibnamefont
  {Covington}}, \bibinfo {author} {\bibfnamefont {M.}~\bibnamefont {Aprili}},
  \bibinfo {author} {\bibfnamefont {E.}~\bibnamefont {Paraoanu}}, \bibinfo
  {author} {\bibfnamefont {L.~H.}\ \bibnamefont {Greene}}, \bibinfo {author}
  {\bibfnamefont {F.}~\bibnamefont {Xu}}, \bibinfo {author} {\bibfnamefont
  {J.}~\bibnamefont {Zhu}},\ and\ \bibinfo {author} {\bibfnamefont {C.~A.}\
  \bibnamefont {Mirkin}},\ }\bibfield  {title} {\bibinfo {title} {{Observation
  of Surface-Induced Broken Time-Reversal Symmetry in
  ${\mathrm{YBa}}_{2}{\mathrm{Cu}}_{3}{O}_{7}$ Tunnel Junctions}},\ }\href
  {https://doi.org/10.1103/PhysRevLett.79.277} {\bibfield  {journal} {\bibinfo
  {journal} {Phys. Rev. Lett.}\ }\textbf {\bibinfo {volume} {79}},\ \bibinfo
  {pages} {277} (\bibinfo {year} {1997})}\BibitemShut {NoStop}%
\bibitem [{\citenamefont {Fogelstr{\ifmmode\ddot{o}\else\"{o}\fi}m}\ \emph
  {et~al.}(1997)\citenamefont {Fogelstr{\ifmmode\ddot{o}\else\"{o}\fi}m},
  \citenamefont {Rainer},\ and\ \citenamefont {Sauls}}]{Fogelstrom1997Jul}%
  \BibitemOpen
  \bibfield  {author} {\bibinfo {author} {\bibfnamefont {M.}~\bibnamefont
  {Fogelstr{\ifmmode\ddot{o}\else\"{o}\fi}m}}, \bibinfo {author} {\bibfnamefont
  {D.}~\bibnamefont {Rainer}},\ and\ \bibinfo {author} {\bibfnamefont {J.~A.}\
  \bibnamefont {Sauls}},\ }\bibfield  {title} {\bibinfo {title} {{Tunneling
  into Current-Carrying Surface States of High- ${\mathit{T}}_{\mathit{c}}$
  Superconductors}},\ }\href {https://doi.org/10.1103/PhysRevLett.79.281}
  {\bibfield  {journal} {\bibinfo  {journal} {Phys. Rev. Lett.}\ }\textbf
  {\bibinfo {volume} {79}},\ \bibinfo {pages} {281} (\bibinfo {year}
  {1997})}\BibitemShut {NoStop}%
\bibitem [{\citenamefont {Kashiwaya}\ and\ \citenamefont
  {Tanaka}(2000)}]{kashiwaya_tunnelling_2000}%
  \BibitemOpen
  \bibfield  {author} {\bibinfo {author} {\bibfnamefont {S.}~\bibnamefont
  {Kashiwaya}}\ and\ \bibinfo {author} {\bibfnamefont {Y.}~\bibnamefont
  {Tanaka}},\ }\bibfield  {title} {\bibinfo {title} {Tunnelling effects on
  surface bound states in unconventional superconductors},\ }\href@noop {}
  {\bibfield  {journal} {\bibinfo  {journal} {Rep. Prog. Phys.}\ }\textbf
  {\bibinfo {volume} {63}},\ \bibinfo {pages} {1641} (\bibinfo {year}
  {2000})}\BibitemShut {NoStop}%
\bibitem [{\citenamefont {Löfwander}\ \emph {et~al.}(2001)\citenamefont
  {Löfwander}, \citenamefont {Shumeiko},\ and\ \citenamefont
  {Wendin}}]{lofwander_andreev_2001}%
  \BibitemOpen
  \bibfield  {author} {\bibinfo {author} {\bibfnamefont {T.}~\bibnamefont
  {Löfwander}}, \bibinfo {author} {\bibfnamefont {V.~S.}\ \bibnamefont
  {Shumeiko}},\ and\ \bibinfo {author} {\bibfnamefont {G.}~\bibnamefont
  {Wendin}},\ }\bibfield  {title} {\bibinfo {title} {{Andreev} bound states in
  high-{$T_\mathrm{c}$} superconducting junctions},\ }\href
  {https://doi.org/10.1088/0953-2048/14/5/201} {\bibfield  {journal} {\bibinfo
  {journal} {Supercond. Sci. Technol.}\ }\textbf {\bibinfo {volume} {14}},\
  \bibinfo {pages} {R53} (\bibinfo {year} {2001})}\BibitemShut {NoStop}%
\bibitem [{\citenamefont {Tsuei}\ and\ \citenamefont
  {Kirtley}(2000)}]{Tsuei2000Oct}%
  \BibitemOpen
  \bibfield  {author} {\bibinfo {author} {\bibfnamefont {C.~C.}\ \bibnamefont
  {Tsuei}}\ and\ \bibinfo {author} {\bibfnamefont {J.~R.}\ \bibnamefont
  {Kirtley}},\ }\bibfield  {title} {\bibinfo {title} {{Pairing symmetry in
  cuprate superconductors}},\ }\href
  {https://doi.org/10.1103/RevModPhys.72.969} {\bibfield  {journal} {\bibinfo
  {journal} {Rev. Mod. Phys.}\ }\textbf {\bibinfo {volume} {72}},\ \bibinfo
  {pages} {969} (\bibinfo {year} {2000})}\BibitemShut {NoStop}%
\bibitem [{\citenamefont {Krupke}\ and\ \citenamefont
  {Deutscher}(1999)}]{Krupke1999}%
  \BibitemOpen
  \bibfield  {author} {\bibinfo {author} {\bibfnamefont {R.}~\bibnamefont
  {Krupke}}\ and\ \bibinfo {author} {\bibfnamefont {G.}~\bibnamefont
  {Deutscher}},\ }\bibfield  {title} {\bibinfo {title} {Anisotropic magnetic
  field dependence of the zero-bias anomaly on in-plane oriented [100]
  ${\mathrm{y}}_{1}{\mathrm{ba}}_{2}{\mathrm{cu}}_{3}{\mathrm{o}}_{7\ensuremath{-}\mathit{x}}/\mathrm{In}$
  tunnel junctions},\ }\href {https://doi.org/10.1103/PhysRevLett.83.4634}
  {\bibfield  {journal} {\bibinfo  {journal} {Phys. Rev. Lett.}\ }\textbf
  {\bibinfo {volume} {83}},\ \bibinfo {pages} {4634} (\bibinfo {year}
  {1999})}\BibitemShut {NoStop}%
\bibitem [{\citenamefont {Wei}\ \emph {et~al.}(1998)\citenamefont {Wei},
  \citenamefont {Yeh}, \citenamefont {Garrigus},\ and\ \citenamefont
  {Strasik}}]{Wei1998}%
  \BibitemOpen
  \bibfield  {author} {\bibinfo {author} {\bibfnamefont {J.~Y.~T.}\
  \bibnamefont {Wei}}, \bibinfo {author} {\bibfnamefont {N.-C.}\ \bibnamefont
  {Yeh}}, \bibinfo {author} {\bibfnamefont {D.~F.}\ \bibnamefont {Garrigus}},\
  and\ \bibinfo {author} {\bibfnamefont {M.}~\bibnamefont {Strasik}},\
  }\bibfield  {title} {\bibinfo {title} {Directional tunneling and andreev
  reflection on
  ${\mathrm{yba}}_{2}{\mathrm{cu}}_{3}{\mathrm{o}}_{7\ensuremath{-}\mathit{\ensuremath{\delta}}}$
  single crystals: Predominance of $\mathit{d}$-wave pairing symmetry verified
  with the generalized blonder, tinkham, and klapwijk theory},\ }\href
  {https://doi.org/10.1103/PhysRevLett.81.2542} {\bibfield  {journal} {\bibinfo
   {journal} {Phys. Rev. Lett.}\ }\textbf {\bibinfo {volume} {81}},\ \bibinfo
  {pages} {2542} (\bibinfo {year} {1998})}\BibitemShut {NoStop}%
\bibitem [{\citenamefont {Dagan}\ and\ \citenamefont
  {Deutscher}(2001)}]{Dagan2001}%
  \BibitemOpen
  \bibfield  {author} {\bibinfo {author} {\bibfnamefont {Y.}~\bibnamefont
  {Dagan}}\ and\ \bibinfo {author} {\bibfnamefont {G.}~\bibnamefont
  {Deutscher}},\ }\bibfield  {title} {\bibinfo {title} {Doping and magnetic
  field dependence of in-plane tunneling into
  ${{\mathrm{YBa}}_{2}{\mathrm{Cu}}_{3}O}_{7\ensuremath{-}\mathit{x}}$:
  Possible evidence for the existence of a quantum critical point},\ }\href
  {https://doi.org/10.1103/PhysRevLett.87.177004} {\bibfield  {journal}
  {\bibinfo  {journal} {Phys. Rev. Lett.}\ }\textbf {\bibinfo {volume} {87}},\
  \bibinfo {pages} {177004} (\bibinfo {year} {2001})}\BibitemShut {NoStop}%
\bibitem [{\citenamefont {Elhalel}\ \emph {et~al.}(2007)\citenamefont
  {Elhalel}, \citenamefont {Beck}, \citenamefont {Leibovitch},\ and\
  \citenamefont {Deutscher}}]{Elhalel2007Mar}%
  \BibitemOpen
  \bibfield  {author} {\bibinfo {author} {\bibfnamefont {G.}~\bibnamefont
  {Elhalel}}, \bibinfo {author} {\bibfnamefont {R.}~\bibnamefont {Beck}},
  \bibinfo {author} {\bibfnamefont {G.}~\bibnamefont {Leibovitch}},\ and\
  \bibinfo {author} {\bibfnamefont {G.}~\bibnamefont {Deutscher}},\ }\bibfield
  {title} {\bibinfo {title} {{Transition from a Mixed to a Pure $d$-Wave
  Symmetry in Superconducting Optimally Doped
  ${\mathrm{YBa}}_{2}{\mathrm{Cu}}_{3}{\mathrm{O}}_{7\ensuremath{-}x}$ Thin
  Films Under Applied Fields}},\ }\href
  {https://doi.org/10.1103/PhysRevLett.98.137002} {\bibfield  {journal}
  {\bibinfo  {journal} {Phys. Rev. Lett.}\ }\textbf {\bibinfo {volume} {98}},\
  \bibinfo {pages} {137002} (\bibinfo {year} {2007})}\BibitemShut {NoStop}%
\bibitem [{\citenamefont {Ngai}\ \emph {et~al.}(2010)\citenamefont {Ngai},
  \citenamefont {Beck}, \citenamefont {Leibovitch}, \citenamefont {Deutscher},\
  and\ \citenamefont {Wei}}]{Ngai2010Aug}%
  \BibitemOpen
  \bibfield  {author} {\bibinfo {author} {\bibfnamefont {J.~H.}\ \bibnamefont
  {Ngai}}, \bibinfo {author} {\bibfnamefont {R.}~\bibnamefont {Beck}}, \bibinfo
  {author} {\bibfnamefont {G.}~\bibnamefont {Leibovitch}}, \bibinfo {author}
  {\bibfnamefont {G.}~\bibnamefont {Deutscher}},\ and\ \bibinfo {author}
  {\bibfnamefont {J.~Y.~T.}\ \bibnamefont {Wei}},\ }\bibfield  {title}
  {\bibinfo {title} {{Local tunneling probe of (110)
  ${\text{Y}}_{0.95}{\text{Ca}}_{0.05}{\text{Ba}}_{2}{\text{Cu}}_{3}{\text{O}}_{7\ensuremath{-}\ensuremath{\delta}}$
  thin films in a magnetic field}},\ }\href
  {https://doi.org/10.1103/PhysRevB.82.054505} {\bibfield  {journal} {\bibinfo
  {journal} {Phys. Rev. B}\ }\textbf {\bibinfo {volume} {82}},\ \bibinfo
  {pages} {054505} (\bibinfo {year} {2010})}\BibitemShut {NoStop}%
\bibitem [{\citenamefont {Matsumoto}\ and\ \citenamefont
  {Shiba}(1995)}]{Matsumoto1995}%
  \BibitemOpen
  \bibfield  {author} {\bibinfo {author} {\bibfnamefont {M.}~\bibnamefont
  {Matsumoto}}\ and\ \bibinfo {author} {\bibfnamefont {H.}~\bibnamefont
  {Shiba}},\ }\bibfield  {title} {\bibinfo {title} {Coexistence of different
  symmetry order parameters near a surface in d-wave superconductors i},\
  }\href {https://doi.org/10.1143/JPSJ.64.3384} {\bibfield  {journal} {\bibinfo
   {journal} {Journal of the Physical Society of Japan}\ }\textbf {\bibinfo
  {volume} {64}},\ \bibinfo {pages} {3384} (\bibinfo {year}
  {1995})}\BibitemShut {NoStop}%
\bibitem [{\citenamefont {Sigrist}(1998)}]{Sigrist1998}%
  \BibitemOpen
  \bibfield  {author} {\bibinfo {author} {\bibfnamefont {M.}~\bibnamefont
  {Sigrist}},\ }\bibfield  {title} {\bibinfo {title} {{Time-Reversal Symmetry
  Breaking States in High-Temperature Superconductors}},\ }\href
  {https://doi.org/10.1143/PTP.99.899} {\bibfield  {journal} {\bibinfo
  {journal} {Progress of Theoretical Physics}\ }\textbf {\bibinfo {volume}
  {99}},\ \bibinfo {pages} {899} (\bibinfo {year} {1998})}\BibitemShut
  {NoStop}%
\bibitem [{\citenamefont {Carmi}\ \emph {et~al.}(2000)\citenamefont {Carmi},
  \citenamefont {Polturak}, \citenamefont {Koren},\ and\ \citenamefont
  {Auerbach}}]{Carmi2000}%
  \BibitemOpen
  \bibfield  {author} {\bibinfo {author} {\bibfnamefont {R.}~\bibnamefont
  {Carmi}}, \bibinfo {author} {\bibfnamefont {E.}~\bibnamefont {Polturak}},
  \bibinfo {author} {\bibfnamefont {G.}~\bibnamefont {Koren}},\ and\ \bibinfo
  {author} {\bibfnamefont {A.}~\bibnamefont {Auerbach}},\ }\bibfield  {title}
  {\bibinfo {title} {{Spontaneous macroscopic magnetization at the
  superconducting transition temperature of YBa$_2$Cu$_3$O$_{7-\delta}$}},\
  }\href {https://doi.org/10.1038/35009062} {\bibfield  {journal} {\bibinfo
  {journal} {Nature}\ }\textbf {\bibinfo {volume} {404}},\ \bibinfo {pages}
  {853} (\bibinfo {year} {2000})}\BibitemShut {NoStop}%
\bibitem [{\citenamefont {Neils}\ and\ \citenamefont
  {Van~Harlingen}(2002)}]{Neils2002}%
  \BibitemOpen
  \bibfield  {author} {\bibinfo {author} {\bibfnamefont {W.~K.}\ \bibnamefont
  {Neils}}\ and\ \bibinfo {author} {\bibfnamefont {D.~J.}\ \bibnamefont
  {Van~Harlingen}},\ }\bibfield  {title} {\bibinfo {title} {Experimental test
  for subdominant superconducting phases with complex order parameters in
  cuprate grain boundary junctions},\ }\href
  {https://doi.org/10.1103/PhysRevLett.88.047001} {\bibfield  {journal}
  {\bibinfo  {journal} {Phys. Rev. Lett.}\ }\textbf {\bibinfo {volume} {88}},\
  \bibinfo {pages} {047001} (\bibinfo {year} {2002})}\BibitemShut {NoStop}%
\bibitem [{\citenamefont {Gustafsson}\ \emph {et~al.}(2013)\citenamefont
  {Gustafsson}, \citenamefont {Golubev}, \citenamefont {Fogelstr\"{o}m},
  \citenamefont {Claeson}, \citenamefont {Kubatkin}, \citenamefont {Bauch},\
  and\ \citenamefont {Lombardi}}]{Gustafsson2013}%
  \BibitemOpen
  \bibfield  {author} {\bibinfo {author} {\bibfnamefont {D.}~\bibnamefont
  {Gustafsson}}, \bibinfo {author} {\bibfnamefont {D.}~\bibnamefont {Golubev}},
  \bibinfo {author} {\bibfnamefont {M.}~\bibnamefont {Fogelstr\"{o}m}},
  \bibinfo {author} {\bibfnamefont {T.}~\bibnamefont {Claeson}}, \bibinfo
  {author} {\bibfnamefont {S.}~\bibnamefont {Kubatkin}}, \bibinfo {author}
  {\bibfnamefont {T.}~\bibnamefont {Bauch}},\ and\ \bibinfo {author}
  {\bibfnamefont {F.}~\bibnamefont {Lombardi}},\ }\bibfield  {title} {\bibinfo
  {title} {{Fully gapped superconductivity in a nanometre-size
  YBa$_2$Cu$_3$O$_{7-\delta}$ island enhanced by a magnetic field}},\ }\href
  {https://doi.org/10.1038/nnano.2012.214} {\bibfield  {journal} {\bibinfo
  {journal} {Nature Nanotechnology}\ }\textbf {\bibinfo {volume} {8}},\
  \bibinfo {pages} {25 } (\bibinfo {year} {2013})}\BibitemShut {NoStop}%
\bibitem [{\citenamefont {Kirtley}\ \emph {et~al.}(2006)\citenamefont
  {Kirtley}, \citenamefont {Tsuei}, \citenamefont {Ariando}, \citenamefont
  {Verwijs}, \citenamefont {Harkema},\ and\ \citenamefont
  {Hilgenkamp}}]{Kirtley2006}%
  \BibitemOpen
  \bibfield  {author} {\bibinfo {author} {\bibfnamefont {J.~R.}\ \bibnamefont
  {Kirtley}}, \bibinfo {author} {\bibfnamefont {C.~C.}\ \bibnamefont {Tsuei}},
  \bibinfo {author} {\bibfnamefont {A.}~\bibnamefont {Ariando}}, \bibinfo
  {author} {\bibfnamefont {C.~J.~M.}\ \bibnamefont {Verwijs}}, \bibinfo
  {author} {\bibfnamefont {S.}~\bibnamefont {Harkema}},\ and\ \bibinfo {author}
  {\bibfnamefont {H.}~\bibnamefont {Hilgenkamp}},\ }\bibfield  {title}
  {\bibinfo {title} {{Angle-resolved phase-sensitive determination of the
  in-plane gap symmetry in YBa$_2$Cu$_3$O$_{7-\delta}$}},\ }\href
  {https://doi.org/10.1038/nphys215} {\bibfield  {journal} {\bibinfo  {journal}
  {Nature Phys.}\ }\textbf {\bibinfo {volume} {2}},\ \bibinfo {pages} {190}
  (\bibinfo {year} {2006})}\BibitemShut {NoStop}%
\bibitem [{\citenamefont {Saadaoui}\ \emph {et~al.}(2011)\citenamefont
  {Saadaoui}, \citenamefont {Morris}, \citenamefont {Salman}, \citenamefont
  {Song}, \citenamefont {Chow}, \citenamefont {Hossain}, \citenamefont {Levy},
  \citenamefont {Parolin}, \citenamefont {Pearson}, \citenamefont {Smadella},
  \citenamefont {Wang}, \citenamefont {Greene}, \citenamefont {Hentges},
  \citenamefont {Kiefl},\ and\ \citenamefont {MacFarlane}}]{Saadaoui2011}%
  \BibitemOpen
  \bibfield  {author} {\bibinfo {author} {\bibfnamefont {H.}~\bibnamefont
  {Saadaoui}}, \bibinfo {author} {\bibfnamefont {G.~D.}\ \bibnamefont
  {Morris}}, \bibinfo {author} {\bibfnamefont {Z.}~\bibnamefont {Salman}},
  \bibinfo {author} {\bibfnamefont {Q.}~\bibnamefont {Song}}, \bibinfo {author}
  {\bibfnamefont {K.~H.}\ \bibnamefont {Chow}}, \bibinfo {author}
  {\bibfnamefont {M.~D.}\ \bibnamefont {Hossain}}, \bibinfo {author}
  {\bibfnamefont {C.~D.~P.}\ \bibnamefont {Levy}}, \bibinfo {author}
  {\bibfnamefont {T.~J.}\ \bibnamefont {Parolin}}, \bibinfo {author}
  {\bibfnamefont {M.~R.}\ \bibnamefont {Pearson}}, \bibinfo {author}
  {\bibfnamefont {M.}~\bibnamefont {Smadella}}, \bibinfo {author}
  {\bibfnamefont {D.}~\bibnamefont {Wang}}, \bibinfo {author} {\bibfnamefont
  {L.~H.}\ \bibnamefont {Greene}}, \bibinfo {author} {\bibfnamefont {P.~J.}\
  \bibnamefont {Hentges}}, \bibinfo {author} {\bibfnamefont {R.~F.}\
  \bibnamefont {Kiefl}},\ and\ \bibinfo {author} {\bibfnamefont {W.~A.}\
  \bibnamefont {MacFarlane}},\ }\bibfield  {title} {\bibinfo {title} {Search
  for broken time-reversal symmetry near the surface of superconducting
  yba${}_{2}$cu${}_{3}$o${}_{7\ensuremath{-}\ensuremath{\delta}}$ films using
  $\ensuremath{\beta}$-detected nuclear magnetic resonance},\ }\href
  {https://doi.org/10.1103/PhysRevB.83.054504} {\bibfield  {journal} {\bibinfo
  {journal} {Phys. Rev. B}\ }\textbf {\bibinfo {volume} {83}},\ \bibinfo
  {pages} {054504} (\bibinfo {year} {2011})}\BibitemShut {NoStop}%
\bibitem [{\citenamefont {H{\aa}kansson}\ \emph {et~al.}(2015)\citenamefont
  {H{\aa}kansson}, \citenamefont {L{\"{o}}fwander},\ and\ \citenamefont
  {Fogelstr{\"{o}}m}}]{Hakansson2015}%
  \BibitemOpen
  \bibfield  {author} {\bibinfo {author} {\bibfnamefont {M.}~\bibnamefont
  {H{\aa}kansson}}, \bibinfo {author} {\bibfnamefont {T.}~\bibnamefont
  {L{\"{o}}fwander}},\ and\ \bibinfo {author} {\bibfnamefont {M.}~\bibnamefont
  {Fogelstr{\"{o}}m}},\ }\bibfield  {title} {\bibinfo {title} {Spontaneously
  broken time-reversal symmetry in high-temperature superconductors},\ }\href
  {https://doi.org/10.1038/nphys3383} {\bibfield  {journal} {\bibinfo
  {journal} {Nature Physics}\ }\textbf {\bibinfo {volume} {11}},\ \bibinfo
  {pages} {755} (\bibinfo {year} {2015})}\BibitemShut {NoStop}%
\bibitem [{\citenamefont {Holmvall}\ \emph {et~al.}(2018)\citenamefont
  {Holmvall}, \citenamefont {Vorontsov}, \citenamefont {Fogelstr{\"{o}}m},\
  and\ \citenamefont {L{\"{o}}fwander}}]{Holmvall2018}%
  \BibitemOpen
  \bibfield  {author} {\bibinfo {author} {\bibfnamefont {P.}~\bibnamefont
  {Holmvall}}, \bibinfo {author} {\bibfnamefont {A.~B.~I.}\ \bibnamefont
  {Vorontsov}}, \bibinfo {author} {\bibfnamefont {M.}~\bibnamefont
  {Fogelstr{\"{o}}m}},\ and\ \bibinfo {author} {\bibfnamefont {T.}~\bibnamefont
  {L{\"{o}}fwander}},\ }\bibfield  {title} {\bibinfo {title} {Broken
  translational symmetry at edges of high-temperature superconductors},\ }\href
  {https://doi.org/10.1038/s41467-018-04531-y} {\bibfield  {journal} {\bibinfo
  {journal} {Nature Communications}\ }\textbf {\bibinfo {volume} {9}},\
  \bibinfo {pages} {2190} (\bibinfo {year} {2018})}\BibitemShut {NoStop}%
\bibitem [{\citenamefont {Holmvall}\ \emph {et~al.}(2020)\citenamefont
  {Holmvall}, \citenamefont {Fogelstr\"om}, \citenamefont {L\"ofwander},\ and\
  \citenamefont {Vorontsov}}]{Holmvall2020}%
  \BibitemOpen
  \bibfield  {author} {\bibinfo {author} {\bibfnamefont {P.}~\bibnamefont
  {Holmvall}}, \bibinfo {author} {\bibfnamefont {M.}~\bibnamefont
  {Fogelstr\"om}}, \bibinfo {author} {\bibfnamefont {T.}~\bibnamefont
  {L\"ofwander}},\ and\ \bibinfo {author} {\bibfnamefont {A.~B.}\ \bibnamefont
  {Vorontsov}},\ }\bibfield  {title} {\bibinfo {title} {Phase crystals},\
  }\href {https://doi.org/10.1103/PhysRevResearch.2.013104} {\bibfield
  {journal} {\bibinfo  {journal} {Phys. Rev. Research}\ }\textbf {\bibinfo
  {volume} {2}},\ \bibinfo {pages} {013104} (\bibinfo {year}
  {2020})}\BibitemShut {NoStop}%
\bibitem [{\citenamefont {Chakraborty}\ \emph {et~al.}(2022)\citenamefont
  {Chakraborty}, \citenamefont {Löfwander}, \citenamefont {Fogelström},\ and\
  \citenamefont {Black-Schaffer}}]{Chakraborty2022}%
  \BibitemOpen
  \bibfield  {author} {\bibinfo {author} {\bibfnamefont {D.}~\bibnamefont
  {Chakraborty}}, \bibinfo {author} {\bibfnamefont {T.}~\bibnamefont
  {Löfwander}}, \bibinfo {author} {\bibfnamefont {M.}~\bibnamefont
  {Fogelström}},\ and\ \bibinfo {author} {\bibfnamefont {A.~M.}\ \bibnamefont
  {Black-Schaffer}},\ }\bibfield  {title} {\bibinfo {title} {Disorder-robust
  phase crystal in high-temperature superconductors stabilized by strong
  correlations},\ }\href {https://doi.org/10.1038/s41535-022-00450-w}
  {\bibfield  {journal} {\bibinfo  {journal} {npj Quantum Materials}\ }\textbf
  {\bibinfo {volume} {7}},\ \bibinfo {pages} {44} (\bibinfo {year}
  {2022})}\BibitemShut {NoStop}%
\bibitem [{\citenamefont {{Honerkamp, C.}}\ \emph {et~al.}(2000)\citenamefont
  {{Honerkamp, C.}}, \citenamefont {{Wakabayashi, K.}},\ and\ \citenamefont
  {{Sigrist, M.}}}]{Honerkamp2000}%
  \BibitemOpen
  \bibfield  {author} {\bibinfo {author} {\bibnamefont {{Honerkamp, C.}}},
  \bibinfo {author} {\bibnamefont {{Wakabayashi, K.}}},\ and\ \bibinfo {author}
  {\bibnamefont {{Sigrist, M.}}},\ }\bibfield  {title} {\bibinfo {title}
  {Instabilities at [110] surfaces of $d_{x^2-y^2}$ superconductors},\ }\href
  {https://doi.org/10.1209/epl/i2000-00280-2} {\bibfield  {journal} {\bibinfo
  {journal} {Europhys. Lett.}\ }\textbf {\bibinfo {volume} {50}},\ \bibinfo
  {pages} {368} (\bibinfo {year} {2000})}\BibitemShut {NoStop}%
\bibitem [{\citenamefont {Potter}\ and\ \citenamefont
  {Lee}(2014)}]{Potter2014}%
  \BibitemOpen
  \bibfield  {author} {\bibinfo {author} {\bibfnamefont {A.~C.}\ \bibnamefont
  {Potter}}\ and\ \bibinfo {author} {\bibfnamefont {P.~A.}\ \bibnamefont
  {Lee}},\ }\bibfield  {title} {\bibinfo {title} {Edge ferromagnetism from
  majorana flat bands: Application to split tunneling-conductance peaks in
  high-${T}_{c}$ cuprate superconductors},\ }\href
  {https://doi.org/10.1103/PhysRevLett.112.117002} {\bibfield  {journal}
  {\bibinfo  {journal} {Phys. Rev. Lett.}\ }\textbf {\bibinfo {volume} {112}},\
  \bibinfo {pages} {117002} (\bibinfo {year} {2014})}\BibitemShut {NoStop}%
\bibitem [{\citenamefont {Clarke}(1972)}]{clarke_experimental_1972}%
  \BibitemOpen
  \bibfield  {author} {\bibinfo {author} {\bibfnamefont {J.}~\bibnamefont
  {Clarke}},\ }\bibfield  {title} {\bibinfo {title} {Experimental {Observation}
  of {Pair}-{Quasiparticle} {Potential} {Difference} in {Nonequilibrium}
  {Superconductors}},\ }\href {https://doi.org/10.1103/PhysRevLett.28.1363}
  {\bibfield  {journal} {\bibinfo  {journal} {Phys. Rev. Lett.}\ }\textbf
  {\bibinfo {volume} {28}},\ \bibinfo {pages} {1363} (\bibinfo {year}
  {1972})}\BibitemShut {NoStop}%
\bibitem [{\citenamefont {Tinkham}\ and\ \citenamefont
  {Clarke}(1972)}]{tinkham_theory_1972}%
  \BibitemOpen
  \bibfield  {author} {\bibinfo {author} {\bibfnamefont {M.}~\bibnamefont
  {Tinkham}}\ and\ \bibinfo {author} {\bibfnamefont {J.}~\bibnamefont
  {Clarke}},\ }\bibfield  {title} {\bibinfo {title} {Theory of
  {Pair}-{Quasiparticle} {Potential} {Difference} in {Nonequilibrium}
  {Superconductors}},\ }\href {https://doi.org/10.1103/PhysRevLett.28.1366}
  {\bibfield  {journal} {\bibinfo  {journal} {Phys. Rev. Lett.}\ }\textbf
  {\bibinfo {volume} {28}},\ \bibinfo {pages} {1366} (\bibinfo {year}
  {1972})}\BibitemShut {NoStop}%
\bibitem [{\citenamefont {Artemenko}\ and\ \citenamefont
  {Volkov}(1979)}]{artemenko_electric_1979}%
  \BibitemOpen
  \bibfield  {author} {\bibinfo {author} {\bibfnamefont {S.~N.}\ \bibnamefont
  {Artemenko}}\ and\ \bibinfo {author} {\bibfnamefont {A.~F.}\ \bibnamefont
  {Volkov}},\ }\bibfield  {title} {\bibinfo {title} {Electric fields and
  collective oscillations in superconductors},\ }\href@noop {} {\bibfield
  {journal} {\bibinfo  {journal} {Sov. Phys. Usp.}\ }\textbf {\bibinfo {volume}
  {22}},\ \bibinfo {pages} {295} (\bibinfo {year} {1979})}\BibitemShut
  {NoStop}%
\bibitem [{\citenamefont {Arutyunov}\ \emph {et~al.}(2018)\citenamefont
  {Arutyunov}, \citenamefont {Chernyaev}, \citenamefont {Karabassov},
  \citenamefont {Lvov}, \citenamefont {Stolyarov},\ and\ \citenamefont
  {Vasenko}}]{arutyunov_relaxation_2018}%
  \BibitemOpen
  \bibfield  {author} {\bibinfo {author} {\bibfnamefont {K.~Y.}\ \bibnamefont
  {Arutyunov}}, \bibinfo {author} {\bibfnamefont {S.~A.}\ \bibnamefont
  {Chernyaev}}, \bibinfo {author} {\bibfnamefont {T.}~\bibnamefont
  {Karabassov}}, \bibinfo {author} {\bibfnamefont {D.~S.}\ \bibnamefont
  {Lvov}}, \bibinfo {author} {\bibfnamefont {V.~S.}\ \bibnamefont
  {Stolyarov}},\ and\ \bibinfo {author} {\bibfnamefont {A.~S.}\ \bibnamefont
  {Vasenko}},\ }\bibfield  {title} {\bibinfo {title}
  {Relaxation of nonequilibrium quasiparticles in mesoscopic size
  superconductors},\ }\href {https://doi.org/10.1088/1361-648X/aad3ea}
  {\bibfield  {journal} {\bibinfo  {journal} {J. Phys. Condens. Matter}\
  }\textbf {\bibinfo {volume} {30}},\ \bibinfo {pages} {343001} (\bibinfo
  {year} {2018})}\BibitemShut {NoStop}%
\bibitem [{\citenamefont {Seja}\ and\ \citenamefont
  {L{\ifmmode\ddot{o}\else\"{o}\fi}fwander}(2021)}]{Seja2021Sep}%
  \BibitemOpen
  \bibfield  {author} {\bibinfo {author} {\bibfnamefont {K.~M.}\ \bibnamefont
  {Seja}}\ and\ \bibinfo {author} {\bibfnamefont {T.}~\bibnamefont
  {L{\ifmmode\ddot{o}\else\"{o}\fi}fwander}},\ }\bibfield  {title} {\bibinfo
  {title} {{Quasiclassical theory of charge transport across mesoscopic
  normal-metal--superconducting heterostructures with current conservation}},\
  }\href {https://doi.org/10.1103/PhysRevB.104.104502} {\bibfield  {journal}
  {\bibinfo  {journal} {Phys. Rev. B}\ }\textbf {\bibinfo {volume} {104}},\
  \bibinfo {pages} {104502} (\bibinfo {year} {2021})}\BibitemShut {NoStop}%
\bibitem [{\citenamefont {Baghdadi}\ \emph {et~al.}(2015)\citenamefont
  {Baghdadi}, \citenamefont {Arpaia}, \citenamefont {Charpentier},
  \citenamefont {Golubev}, \citenamefont {Bauch},\ and\ \citenamefont
  {Lombardi}}]{Baghdadi2015Jul}%
  \BibitemOpen
  \bibfield  {author} {\bibinfo {author} {\bibfnamefont {R.}~\bibnamefont
  {Baghdadi}}, \bibinfo {author} {\bibfnamefont {R.}~\bibnamefont {Arpaia}},
  \bibinfo {author} {\bibfnamefont {S.}~\bibnamefont {Charpentier}}, \bibinfo
  {author} {\bibfnamefont {D.}~\bibnamefont {Golubev}}, \bibinfo {author}
  {\bibfnamefont {T.}~\bibnamefont {Bauch}},\ and\ \bibinfo {author}
  {\bibfnamefont {F.}~\bibnamefont {Lombardi}},\ }\bibfield  {title} {\bibinfo
  {title} {{Fabricating Nanogaps in
  ${\mathrm{YBa}}_{2}{\mathrm{Cu}}_{3}{\mathrm{O}}_{7\ensuremath{-}\ensuremath{\delta}}$
  for Hybrid Proximity-Based Josephson Junctions}},\ }\href
  {https://doi.org/10.1103/PhysRevApplied.4.014022} {\bibfield  {journal}
  {\bibinfo  {journal} {Phys. Rev. Appl.}\ }\textbf {\bibinfo {volume} {4}},\
  \bibinfo {pages} {014022} (\bibinfo {year} {2015})}\BibitemShut {NoStop}%
\bibitem [{\citenamefont {Baghdadi}\ \emph {et~al.}(2017)\citenamefont
  {Baghdadi}, \citenamefont {Abay}, \citenamefont {Golubev}, \citenamefont
  {Bauch},\ and\ \citenamefont {Lombardi}}]{Baghdadi2017}%
  \BibitemOpen
  \bibfield  {author} {\bibinfo {author} {\bibfnamefont {R.}~\bibnamefont
  {Baghdadi}}, \bibinfo {author} {\bibfnamefont {S.}~\bibnamefont {Abay}},
  \bibinfo {author} {\bibfnamefont {D.}~\bibnamefont {Golubev}}, \bibinfo
  {author} {\bibfnamefont {T.}~\bibnamefont {Bauch}},\ and\ \bibinfo {author}
  {\bibfnamefont {F.}~\bibnamefont {Lombardi}},\ }\bibfield  {title} {\bibinfo
  {title} {Josephson effect through
  ${\mathrm{yba}}_{2}{\mathrm{cu}}_{3}{\mathrm{o}}_{7\ensuremath{-}\ensuremath{\delta}}$/au-encapsulated
  nanogaps},\ }\href {https://doi.org/10.1103/PhysRevB.95.174510} {\bibfield
  {journal} {\bibinfo  {journal} {Phys. Rev. B}\ }\textbf {\bibinfo {volume}
  {95}},\ \bibinfo {pages} {174510} (\bibinfo {year} {2017})}\BibitemShut
  {NoStop}%
\bibitem [{\citenamefont {Seja}\ \emph {et~al.}(2022)\citenamefont {Seja},
  \citenamefont {Jacob},\ and\ \citenamefont {L\"ofwander}}]{Seja2022Mar}%
  \BibitemOpen
  \bibfield  {author} {\bibinfo {author} {\bibfnamefont {K.~M.}\ \bibnamefont
  {Seja}}, \bibinfo {author} {\bibfnamefont {L.}~\bibnamefont {Jacob}},\ and\
  \bibinfo {author} {\bibfnamefont {T.}~\bibnamefont {L\"ofwander}},\
  }\bibfield  {title} {\bibinfo {title} {Thermopower and thermophase in a
  $d$-wave superconductor},\ }\href
  {https://doi.org/10.1103/PhysRevB.105.104506} {\bibfield  {journal} {\bibinfo
   {journal} {Phys. Rev. B}\ }\textbf {\bibinfo {volume} {105}},\ \bibinfo
  {pages} {104506} (\bibinfo {year} {2022})}\BibitemShut {NoStop}%
\bibitem [{\citenamefont {Eilenberger}(1968)}]{Eilenberger1968Apr}%
  \BibitemOpen
  \bibfield  {author} {\bibinfo {author} {\bibfnamefont {G.}~\bibnamefont
  {Eilenberger}},\ }\bibfield  {title} {\bibinfo {title} {{Transformation of
  Gorkov's equation for type II superconductors into transport-like
  equations}},\ }\href {https://doi.org/10.1007/BF01379803} {\bibfield
  {journal} {\bibinfo  {journal} {Z. Phys. A: Hadrons Nucl.}\ }\textbf
  {\bibinfo {volume} {214}},\ \bibinfo {pages} {195} (\bibinfo {year}
  {1968})}\BibitemShut {NoStop}%
\bibitem [{\citenamefont {Larkin}\ and\ \citenamefont
  {Ovchinnikov}(1968)}]{Larkin1969}%
  \BibitemOpen
  \bibfield  {author} {\bibinfo {author} {\bibfnamefont {A.}~\bibnamefont
  {Larkin}}\ and\ \bibinfo {author} {\bibfnamefont {Y.~N.}\ \bibnamefont
  {Ovchinnikov}},\ }\bibfield  {title} {\bibinfo {title} {{Quasiclassical
  Method in the Theory of Superconductivity}},\ }\href@noop {} {\bibfield
  {journal} {\bibinfo  {journal} {ZhETF}\ }\textbf {\bibinfo {volume} {55}},\
  \bibinfo {pages} {2262} (\bibinfo {year} {1968})}\BibitemShut {NoStop}%
\bibitem [{\citenamefont {Eliashberg}(1971)}]{Eliashberg1971}%
  \BibitemOpen
  \bibfield  {author} {\bibinfo {author} {\bibfnamefont {G.}~\bibnamefont
  {Eliashberg}},\ }\bibfield  {title} {\bibinfo {title} {Inelastic electron
  collisions and nonequilibrium stationary states in superconductors [eng.:
  Sov. phys. jetp 34, 668–676 (1972)]},\ }\href
  {http://jetp.ras.ru/cgi-bin/e/index/e/34/3/p668?a=list} {\bibfield  {journal}
  {\bibinfo  {journal} {Zh. Eksp. i Teor. Fiz.}\ }\textbf {\bibinfo {volume}
  {61}},\ \bibinfo {pages} {1254} (\bibinfo {year} {1971})}\BibitemShut
  {NoStop}%
\bibitem [{\citenamefont {Abrikosov}\ \emph {et~al.}(1975)\citenamefont
  {Abrikosov}, \citenamefont {Gorkov},\ and\ \citenamefont
  {Dzyaloshinski}}]{AGD:Book}%
  \BibitemOpen
  \bibfield  {author} {\bibinfo {author} {\bibfnamefont {A.~A.}\ \bibnamefont
  {Abrikosov}}, \bibinfo {author} {\bibfnamefont {L.~P.}\ \bibnamefont
  {Gorkov}},\ and\ \bibinfo {author} {\bibfnamefont {I.~E.}\ \bibnamefont
  {Dzyaloshinski}},\ }\href@noop {} {\emph {\bibinfo {title} {Methods of
  quantum field theory in statistical physics}}}\ (\bibinfo  {publisher} {Dover
  Publications, Inc.},\ \bibinfo {address} {New York},\ \bibinfo {year}
  {1975})\BibitemShut {NoStop}%
\bibitem [{\citenamefont {Arfi}\ \emph {et~al.}(1988)\citenamefont {Arfi},
  \citenamefont {Bahlouli}, \citenamefont {Pethick},\ and\ \citenamefont
  {Pines}}]{arf88}%
  \BibitemOpen
  \bibfield  {author} {\bibinfo {author} {\bibfnamefont {B.}~\bibnamefont
  {Arfi}}, \bibinfo {author} {\bibfnamefont {H.}~\bibnamefont {Bahlouli}},
  \bibinfo {author} {\bibfnamefont {C.~J.}\ \bibnamefont {Pethick}},\ and\
  \bibinfo {author} {\bibfnamefont {D.}~\bibnamefont {Pines}},\ }\bibfield
  {title} {\bibinfo {title} {Unusual transport effects in anisotropic
  superconductors},\ }\href {https://doi.org/10.1103/PhysRevLett.60.2206}
  {\bibfield  {journal} {\bibinfo  {journal} {Phys. Rev. Lett.}\ }\textbf
  {\bibinfo {volume} {60}},\ \bibinfo {pages} {2206} (\bibinfo {year}
  {1988})}\BibitemShut {NoStop}%
\bibitem [{\citenamefont {Xu}\ \emph {et~al.}(1995)\citenamefont {Xu},
  \citenamefont {Yip},\ and\ \citenamefont {Sauls}}]{xu95}%
  \BibitemOpen
  \bibfield  {author} {\bibinfo {author} {\bibfnamefont {D.}~\bibnamefont
  {Xu}}, \bibinfo {author} {\bibfnamefont {S.~K.}\ \bibnamefont {Yip}},\ and\
  \bibinfo {author} {\bibfnamefont {J.~A.}\ \bibnamefont {Sauls}},\ }\bibfield
  {title} {\bibinfo {title} {Nonlinear {Meissner} effect in unconventional
  superconductors},\ }\href {https://doi.org/10.1103/PhysRevB.51.16233}
  {\bibfield  {journal} {\bibinfo  {journal} {Phys. Rev. B}\ }\textbf {\bibinfo
  {volume} {51}},\ \bibinfo {pages} {16233} (\bibinfo {year}
  {1995})}\BibitemShut {NoStop}%
\bibitem [{\citenamefont {Graf}\ \emph {et~al.}(1996)\citenamefont {Graf},
  \citenamefont {Yip}, \citenamefont {Sauls},\ and\ \citenamefont
  {Rainer}}]{graf96}%
  \BibitemOpen
  \bibfield  {author} {\bibinfo {author} {\bibfnamefont {M.~J.}\ \bibnamefont
  {Graf}}, \bibinfo {author} {\bibfnamefont {S.-K.}\ \bibnamefont {Yip}},
  \bibinfo {author} {\bibfnamefont {J.~A.}\ \bibnamefont {Sauls}},\ and\
  \bibinfo {author} {\bibfnamefont {D.}~\bibnamefont {Rainer}},\ }\bibfield
  {title} {\bibinfo {title} {Electronic thermal conductivity and the
  {Wiedemann-Franz} law for unconventional superconductors},\ }\href
  {https://doi.org/10.1103/PhysRevB.53.15147} {\bibfield  {journal} {\bibinfo
  {journal} {Phys. Rev. B}\ }\textbf {\bibinfo {volume} {53}},\ \bibinfo
  {pages} {15147} (\bibinfo {year} {1996})}\BibitemShut {NoStop}%
\bibitem [{\citenamefont {Poenicke}\ \emph {et~al.}(1999)\citenamefont
  {Poenicke}, \citenamefont {Barash}, \citenamefont {Bruder},\ and\
  \citenamefont {Istyukov}}]{Poenicke1999}%
  \BibitemOpen
  \bibfield  {author} {\bibinfo {author} {\bibfnamefont {A.}~\bibnamefont
  {Poenicke}}, \bibinfo {author} {\bibfnamefont {Y.~S.}\ \bibnamefont
  {Barash}}, \bibinfo {author} {\bibfnamefont {C.}~\bibnamefont {Bruder}},\
  and\ \bibinfo {author} {\bibfnamefont {V.}~\bibnamefont {Istyukov}},\
  }\bibfield  {title} {\bibinfo {title} {Broadening of andreev bound states in
  ${d}_{{x}^{2}\ensuremath{-}{y}^{2}}$ superconductors},\ }\href
  {https://doi.org/10.1103/PhysRevB.59.7102} {\bibfield  {journal} {\bibinfo
  {journal} {Phys. Rev. B}\ }\textbf {\bibinfo {volume} {59}},\ \bibinfo
  {pages} {7102} (\bibinfo {year} {1999})}\BibitemShut {NoStop}%
\bibitem [{\citenamefont {L\"ofwander}(2004)}]{LofwanderProximity2004}%
  \BibitemOpen
  \bibfield  {author} {\bibinfo {author} {\bibfnamefont {T.}~\bibnamefont
  {L\"ofwander}},\ }\bibfield  {title} {\bibinfo {title} {Proximity effect in
  normal metal--high-${T}_{c}$ superconductor contacts},\ }\href
  {https://doi.org/10.1103/PhysRevB.70.094518} {\bibfield  {journal} {\bibinfo
  {journal} {Phys. Rev. B}\ }\textbf {\bibinfo {volume} {70}},\ \bibinfo
  {pages} {094518} (\bibinfo {year} {2004})}\BibitemShut {NoStop}%
\bibitem [{\citenamefont {L{\ifmmode\ddot{o}\else\"{o}\fi}fwander}\ and\
  \citenamefont
  {Fogelstr{\ifmmode\ddot{o}\else\"{o}\fi}m}(2004)}]{Lofwander2004Jul}%
  \BibitemOpen
  \bibfield  {author} {\bibinfo {author} {\bibfnamefont {T.}~\bibnamefont
  {L{\ifmmode\ddot{o}\else\"{o}\fi}fwander}}\ and\ \bibinfo {author}
  {\bibfnamefont {M.}~\bibnamefont
  {Fogelstr{\ifmmode\ddot{o}\else\"{o}\fi}m}},\ }\bibfield  {title} {\bibinfo
  {title} {{Large thermoelectric effects in unconventional superconductors}},\
  }\href {https://doi.org/10.1103/PhysRevB.70.024515} {\bibfield  {journal}
  {\bibinfo  {journal} {Phys. Rev. B}\ }\textbf {\bibinfo {volume} {70}},\
  \bibinfo {pages} {024515} (\bibinfo {year} {2004})}\BibitemShut {NoStop}%
\bibitem [{\citenamefont {Eschrig}\ \emph {et~al.}(1999)\citenamefont
  {Eschrig}, \citenamefont {Sauls},\ and\ \citenamefont
  {Rainer}}]{Eschrig1999Oct}%
  \BibitemOpen
  \bibfield  {author} {\bibinfo {author} {\bibfnamefont {M.}~\bibnamefont
  {Eschrig}}, \bibinfo {author} {\bibfnamefont {J.~A.}\ \bibnamefont {Sauls}},\
  and\ \bibinfo {author} {\bibfnamefont {D.}~\bibnamefont {Rainer}},\
  }\bibfield  {title} {\bibinfo {title} {{Electromagnetic response of a vortex
  in layered superconductors}},\ }\href
  {https://doi.org/10.1103/PhysRevB.60.10447} {\bibfield  {journal} {\bibinfo
  {journal} {Phys. Rev. B}\ }\textbf {\bibinfo {volume} {60}},\ \bibinfo
  {pages} {10447} (\bibinfo {year} {1999})}\BibitemShut {NoStop}%
\bibitem [{\citenamefont {Sánchez-Cañizares}\ and\ \citenamefont
  {Sols}(1997)}]{sanchez-canizares_self-consistent_1997}%
  \BibitemOpen
  \bibfield  {author} {\bibinfo {author} {\bibfnamefont {J.}~\bibnamefont
  {Sánchez-Cañizares}}\ and\ \bibinfo {author} {\bibfnamefont
  {F.}~\bibnamefont {Sols}},\ }\bibfield  {title} {\bibinfo {title}
  {Self-consistent scattering description of transport in normal-superconductor
  structures},\ }\href@noop {} {\bibfield  {journal} {\bibinfo  {journal}
  {Phys. Rev. B}\ }\textbf {\bibinfo {volume} {55}},\ \bibinfo {pages} {531}
  (\bibinfo {year} {1997})}\BibitemShut {NoStop}%
\bibitem [{\citenamefont {Keizer}\ \emph {et~al.}(2006)\citenamefont {Keizer},
  \citenamefont {Flokstra}, \citenamefont {Aarts},\ and\ \citenamefont
  {Klapwijk}}]{Keizer2006Apr}%
  \BibitemOpen
  \bibfield  {author} {\bibinfo {author} {\bibfnamefont {R.~S.}\ \bibnamefont
  {Keizer}}, \bibinfo {author} {\bibfnamefont {M.~G.}\ \bibnamefont
  {Flokstra}}, \bibinfo {author} {\bibfnamefont {J.}~\bibnamefont {Aarts}},\
  and\ \bibinfo {author} {\bibfnamefont {T.~M.}\ \bibnamefont {Klapwijk}},\
  }\bibfield  {title} {\bibinfo {title} {{Critical Voltage of a Mesoscopic
  Superconductor}},\ }\href {https://doi.org/10.1103/PhysRevLett.96.147002}
  {\bibfield  {journal} {\bibinfo  {journal} {Phys. Rev. Lett.}\ }\textbf
  {\bibinfo {volume} {96}},\ \bibinfo {pages} {147002} (\bibinfo {year}
  {2006})}\BibitemShut {NoStop}%
\bibitem [{\citenamefont {Bardeen}(1962)}]{Bardeen1962}%
  \BibitemOpen
  \bibfield  {author} {\bibinfo {author} {\bibfnamefont {J.}~\bibnamefont
  {Bardeen}},\ }\bibfield  {title} {\bibinfo {title} {Critical fields and
  currents in superconductors},\ }\href
  {https://doi.org/10.1103/RevModPhys.34.667} {\bibfield  {journal} {\bibinfo
  {journal} {Rev. Mod. Phys.}\ }\textbf {\bibinfo {volume} {34}},\ \bibinfo
  {pages} {667} (\bibinfo {year} {1962})}\BibitemShut {NoStop}%
\bibitem [{\citenamefont {Lee}(1993)}]{lee93}%
  \BibitemOpen
  \bibfield  {author} {\bibinfo {author} {\bibfnamefont {P.~A.}\ \bibnamefont
  {Lee}},\ }\bibfield  {title} {\bibinfo {title} {Localized states in a d-wave
  superconductor},\ }\href {https://doi.org/10.1103/PhysRevLett.71.1887}
  {\bibfield  {journal} {\bibinfo  {journal} {Phys. Rev. Lett.}\ }\textbf
  {\bibinfo {volume} {71}},\ \bibinfo {pages} {1887} (\bibinfo {year}
  {1993})}\BibitemShut {NoStop}%
\bibitem [{\citenamefont {L\"ofwander}\ \emph {et~al.}(2003)\citenamefont
  {L\"ofwander}, \citenamefont {Fogelstr\"om},\ and\ \citenamefont
  {Sauls}}]{Lofwander2003}%
  \BibitemOpen
  \bibfield  {author} {\bibinfo {author} {\bibfnamefont {T.}~\bibnamefont
  {L\"ofwander}}, \bibinfo {author} {\bibfnamefont {M.}~\bibnamefont
  {Fogelstr\"om}},\ and\ \bibinfo {author} {\bibfnamefont {J.~A.}\ \bibnamefont
  {Sauls}},\ }\bibfield  {title} {\bibinfo {title} {Shot noise in normal
  metal--d-wave superconducting junctions},\ }\href
  {https://doi.org/10.1103/PhysRevB.68.054504} {\bibfield  {journal} {\bibinfo
  {journal} {Phys. Rev. B}\ }\textbf {\bibinfo {volume} {68}},\ \bibinfo
  {pages} {054504} (\bibinfo {year} {2003})}\BibitemShut {NoStop}%
\bibitem [{\citenamefont {Nagato}\ \emph {et~al.}(1993)\citenamefont {Nagato},
  \citenamefont {Nagai},\ and\ \citenamefont {Hara}}]{Nagato1993}%
  \BibitemOpen
  \bibfield  {author} {\bibinfo {author} {\bibfnamefont {Y.}~\bibnamefont
  {Nagato}}, \bibinfo {author} {\bibfnamefont {K.}~\bibnamefont {Nagai}},\ and\
  \bibinfo {author} {\bibfnamefont {J.}~\bibnamefont {Hara}},\ }\bibfield
  {title} {\bibinfo {title} {{Theory of the Andreev reflection and the density
  of states in proximity contact normal-superconducting infinite
  double-layer}},\ }\href {https://doi.org/10.1007/BF00682280} {\bibfield
  {journal} {\bibinfo  {journal} {J. Low Temp. Phys.}\ }\textbf {\bibinfo
  {volume} {93}},\ \bibinfo {pages} {33} (\bibinfo {year} {1993})}\BibitemShut
  {NoStop}%
\bibitem [{\citenamefont {Schopohl}\ and\ \citenamefont
  {Maki}(1995)}]{Schopohl1995}%
  \BibitemOpen
  \bibfield  {author} {\bibinfo {author} {\bibfnamefont {N.}~\bibnamefont
  {Schopohl}}\ and\ \bibinfo {author} {\bibfnamefont {K.}~\bibnamefont
  {Maki}},\ }\bibfield  {title} {\bibinfo {title} {{Quasiparticle spectrum
  around a vortex line in a d-wave superconductor}},\ }\href
  {https://doi.org/10.1103/PhysRevB.52.490} {\bibfield  {journal} {\bibinfo
  {journal} {Phys. Rev. B}\ }\textbf {\bibinfo {volume} {52}},\ \bibinfo
  {pages} {490} (\bibinfo {year} {1995})}\BibitemShut {NoStop}%
\bibitem [{\citenamefont {Schopohl}(1998)}]{Schopohl1998}%
  \BibitemOpen
  \bibfield  {author} {\bibinfo {author} {\bibfnamefont {N.}~\bibnamefont
  {Schopohl}},\ }\bibfield  {title} {\bibinfo {title} {{Transformation of the
  Eilenberger Equations of Superconductivity to a Scalar Riccati Equation}},\
  }\href {https://arxiv.org/abs/cond-mat/9804064v1} {\bibfield  {journal}
  {\bibinfo  {journal} {arXiv}\ } (\bibinfo {year} {1998})},\ \Eprint
  {https://arxiv.org/abs/cond-mat/9804064} {cond-mat/9804064} \BibitemShut
  {NoStop}%
\bibitem [{\citenamefont {Eschrig}(2000)}]{eschrig_distribution_2000}%
  \BibitemOpen
  \bibfield  {author} {\bibinfo {author} {\bibfnamefont {M.}~\bibnamefont
  {Eschrig}},\ }\bibfield  {title} {\bibinfo {title} {Distribution functions in
  nonequilibrium theory of superconductivity and {Andreev} spectroscopy in
  unconventional superconductors},\ }\href
  {https://doi.org/10.1103/PhysRevB.61.9061} {\bibfield  {journal} {\bibinfo
  {journal} {Phys. Rev. B}\ }\textbf {\bibinfo {volume} {61}},\ \bibinfo
  {pages} {9061} (\bibinfo {year} {2000})}\BibitemShut {NoStop}%
\bibitem [{\citenamefont {Eschrig}(2009)}]{Eschrig2009Oct}%
  \BibitemOpen
  \bibfield  {author} {\bibinfo {author} {\bibfnamefont {M.}~\bibnamefont
  {Eschrig}},\ }\bibfield  {title} {\bibinfo {title} {{Scattering problem in
  nonequilibrium quasiclassical theory of metals and superconductors: General
  boundary conditions and applications}},\ }\href
  {https://doi.org/10.1103/PhysRevB.80.134511} {\bibfield  {journal} {\bibinfo
  {journal} {Phys. Rev. B}\ }\textbf {\bibinfo {volume} {80}},\ \bibinfo
  {pages} {134511} (\bibinfo {year} {2009})}\BibitemShut {NoStop}%
\bibitem [{\citenamefont {Zhao}\ \emph {et~al.}(2004)\citenamefont {Zhao},
  \citenamefont {L{\ifmmode\ddot{o}\else\"{o}\fi}fwander},\ and\ \citenamefont
  {Sauls}}]{Zhao2004Oct}%
  \BibitemOpen
  \bibfield  {author} {\bibinfo {author} {\bibfnamefont {E.}~\bibnamefont
  {Zhao}}, \bibinfo {author} {\bibfnamefont {T.}~\bibnamefont
  {L{\ifmmode\ddot{o}\else\"{o}\fi}fwander}},\ and\ \bibinfo {author}
  {\bibfnamefont {J.~A.}\ \bibnamefont {Sauls}},\ }\bibfield  {title} {\bibinfo
  {title} {{Nonequilibrium superconductivity near spin-active interfaces}},\
  }\href {https://doi.org/10.1103/PhysRevB.70.134510} {\bibfield  {journal}
  {\bibinfo  {journal} {Phys. Rev. B}\ }\textbf {\bibinfo {volume} {70}},\
  \bibinfo {pages} {134510} (\bibinfo {year} {2004})}\BibitemShut {NoStop}%
\bibitem [{\citenamefont {Schmid}\ and\ \citenamefont
  {Sch{\ifmmode\ddot{o}\else\"{o}\fi}n}(1975)}]{Schmid1975Jul}%
  \BibitemOpen
  \bibfield  {author} {\bibinfo {author} {\bibfnamefont {A.}~\bibnamefont
  {Schmid}}\ and\ \bibinfo {author} {\bibfnamefont {G.}~\bibnamefont
  {Sch{\ifmmode\ddot{o}\else\"{o}\fi}n}},\ }\bibfield  {title} {\bibinfo
  {title} {{Linearized kinetic equations and relaxation processes of a
  superconductor near {T$_c$}}},\ }\href {https://doi.org/10.1007/BF00115264}
  {\bibfield  {journal} {\bibinfo  {journal} {J. Low Temp. Phys.}\ }\textbf
  {\bibinfo {volume} {20}},\ \bibinfo {pages} {207} (\bibinfo {year}
  {1975})}\BibitemShut {NoStop}%
\bibitem [{\citenamefont {Datta}(2017)}]{Datta2017Feb}%
  \BibitemOpen
  \bibfield  {author} {\bibinfo {author} {\bibfnamefont {S.}~\bibnamefont
  {Datta}},\ }\href {https://doi.org/10.1142/10440} {\emph {\bibinfo {title}
  {{Lessons from Nanoelectronics {$\vert$} Lessons from Nanoscience: A Lecture
  Notes Series}}}},\ Vol.~\bibinfo {volume} {5}\ (\bibinfo  {publisher} {World
  Scientific Publishing Company},\ \bibinfo {year} {2017})\BibitemShut
  {NoStop}%
\end{thebibliography}

\providecommand{\noopsort}[1]{}\providecommand{\singleletter}[1]{#1}%

\end{document}